\pdfoutput=1
\documentclass[11pt]{article}

\usepackage[final]{acl}

\usepackage{times}
\usepackage{latexsym}
\usepackage{fdsymbol}

\usepackage[T1]{fontenc}
\usepackage[utf8]{inputenc}

\usepackage{microtype}

\usepackage{inconsolata}

\usepackage{amsmath}
\usepackage{amssymb}
\usepackage{mathtools}
\usepackage{amsthm}

\usepackage{graphicx}
\usepackage{microtype}
\usepackage{subfigure}
\usepackage{booktabs} % for professional tables
\usepackage{multirow}  % 支持单元格合并
\usepackage{makecell} % For multi-line cells and formatting
\usepackage{tabularx} % 用于设置表格总宽度
\usepackage{array}     % 支持 m{} 列类型
\usepackage{adjustbox}
\usepackage{tcolorbox}
\usepackage{xcolor}
\usepackage{colortbl}    % 定义 \rowcolor
\usepackage{pifont}
\def\halfcheckmark{\ding{52}\rotatebox[origin=c]{-9.2}{\kern-0.7em\ding{55}}}

\definecolor{redcolor}{HTML}{d84152}

\usepackage{xspace}

\newtheorem*{remark}{Remark}

\definecolor{customcyan}{RGB}{0, 158, 115} 
\definecolor{tealblue}{RGB}{0, 114, 178}
\definecolor{darkorange}{RGB}{213, 94, 0} 

\usepackage[textsize=tiny]{todonotes}

\title{LocAgent: Graph-Guided LLM Agents for Code Localization}

\author{
Zhaoling Chen$^{\spadesuit *}$, 
Xiangru Tang$^{\spadesuit}\thanks{\quad Equal contribution. This work was done during Zhaoling's time at Yale.}$, 
Gangda Deng$^{\vardiamondsuit *}$,
Fang Wu$^{\clubsuit}$,
Jialong Wu$^{\spadesuit}$,
Zhiwei Jiang,\\
\textbf{Viktor Prasanna$^{\vardiamondsuit}$,}
\textbf{Arman Cohan$^{\spadesuit}$,}
\textbf{Xingyao Wang$^{\varheartsuit}$}
\\
$^\spadesuit$Yale University \quad
$^\vardiamondsuit$University of Southern California \quad
$^\clubsuit$Stanford University \quad
$^\varheartsuit$All Hands AI\\
\texttt{xiangru.tang@yale.edu}, \texttt{gangdade@usc.edu}, \texttt{xingyao@all-hands.dev}
}
\begin{document}

\maketitle

\begin{abstract}
% Lack of motivation of why we need to model codebase as graph. 
% Given an issue description, code localization aims to pinpoint the underlying affected code regions.

%Code localization is critical for various software development tasks, including automated bug fixing, refactoring, and feature additions. 
Code localization--identifying precisely where in a codebase changes need to be made--is a fundamental yet challenging task in software maintenance. 
Existing approaches struggle to efficiently navigate complex codebases when identifying relevant code sections.
The challenge lies in bridging natural language problem descriptions with the appropriate code elements, often requiring reasoning across hierarchical structures and multiple dependencies.
We introduce \textsc{LocAgent}, a framework that addresses code localization through graph-based representation. 
By parsing codebases into directed heterogeneous graphs, \textsc{LocAgent} creates a lightweight representation that captures code structures (files, classes, functions) and their dependencies (imports, invocations, inheritance), enabling LLM agents to effectively search and locate relevant entities through powerful multi-hop reasoning.
Experimental results on real-world benchmarks demonstrate that our approach significantly enhances accuracy in code localization.
Notably, our method with the fine-tuned \texttt{Qwen-2.5-Coder-Instruct-32B} model achieves comparable results to SOTA proprietary models at greatly reduced cost (approximately 86\% reduction), reaching up to 92.7\% accuracy on file-level localization while improving downstream GitHub issue resolution success rates by 12\% for multiple attempts (Pass@10). Our code is available at \url{https://github.com/gersteinlab/LocAgent}.

\end{abstract}
% \footnote{\texttt{Claude-3-5-sonnet-20241022}}
\begin{figure}[t!]  % 使用 figure* 环境
    \includegraphics[width=\linewidth, trim=42 270 465 88, clip]{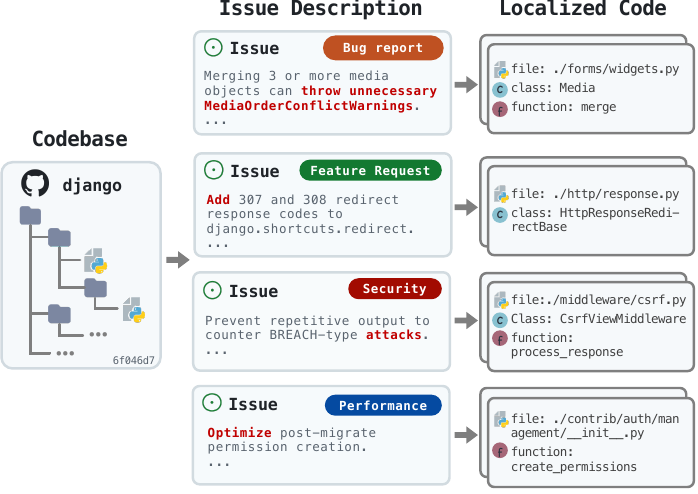}  % 使用 \textwidth 占满整个宽度
    \caption{Code localization across four common programming scenarios. Given a codebase and an issue description, the goal of code localization is to identify the relevant code snippets that require modification to resolve the issue.}
    \label{fig:figure_1}
\vspace{-.6cm}
\end{figure}

\section{Introduction}
Code localization can be viewed as an information retrieval (IR) task that aims to identify relevant code snippets given natural language descriptions \cite{yu2025orcaloca,yang2024sweagent,xia2024agentless}. 
Developers spend up to 66\% of their debugging time \cite{bohme2017bug} understanding code to make changes, and automated tools often struggle with the same challenge. Poor code localization leads to incomplete fixes, introduces new bugs, and significantly extends development cycles. 
Unlike traditional retrieval tasks that primarily focus on lexical or semantic matching between queries and documents \cite{guo2016deep,guo2020deep}, code localization requires bridging the gap between natural language and programming languages. It also necessitates reasoning capabilities to analyze the issue, while considering the structural and semantic properties of code~\cite{lewis2020retrieval,guu2020retrieval,qu2020open}. This capability has become fundamental to powerful AI assistants~\cite{openai2023chatgpt, anthropic2023claude}, code-aware search engines~\cite{perplexityai2023}, and automated programming agents~\cite{cognitionai2024devin, wang2024openhands, aider2024swe}. In particular, accurate code localization is crucial for software maintenance and evolution, as it enables precise code modifications for bug fixes, refactoring, and feature additions~\cite{wang2024coderagbench}, thereby streamlining the development workflow.

Existing approaches to code localization face significant limitations. Dense retrieval methods require maintaining and continuously updating vector representations of the entire codebase~\cite{wang2023codet5opencodelarge,günther2023jinaembeddingsnovelset}, creating engineering challenges for large, evolving repositories where code changes frequently. While LLMs demonstrate strong code understanding capabilities~\cite{kang2023preliminary,wu2023large}, models with large context windows cannot process entire codebases at once, necessitating strategic navigation through relevant parts.
Moreover, issue descriptions often mention only symptoms rather than underlying causes. For instance, a report of `XSS vulnerability in user profile' might require changes to a shared validation utility used throughout the codebase but not explicitly referenced in the issue. This disconnect between issue descriptions and affected code components presents a substantial challenge for traditional retrieval approaches, which struggle to trace implicit dependencies across the codebase structure.
Recent agent-based methods attempt to address these limitations through iterative exploration~\cite{yang2024sweagent,qin2024agentfl} but still struggle to efficiently navigate and comprehend complex code structures and dependencies, particularly when multi-hop reasoning is required to trace from issue descriptions to affected code regions that aren't directly mentioned.

This raises a key question: \textit{How can we design efficient indexing as intermediate representations that are structure-aware and both easy and performant for LLM agents to consume?}
It is intuitive to design an agentic retrieval system that carefully combines traditional IR methods and LLM agent's reasoning ability to achieve accurate, efficient, and cost-effective code localization in codebases.

To address this challenge, we propose \textsc{LocAgent}, a framework that builds directed heterogeneous graph indexing to unify code structures, dependencies, and contents. Our approach leverages a structured graph representation that enables powerful multi-hop reasoning capabilities, allowing agents to navigate complex dependency relationships between code elements even when target code isn't explicitly mentioned in issue descriptions. This graph-based approach significantly outperforms previous methods on challenging localization tasks that require traversing multiple code relationships. Our lightweight representation, coupled with sparse indexing techniques, enables efficient entity search while maintaining rich structural information. The indexing process typically takes only a few seconds per codebase, making it highly practical for real-time use. The framework integrates a set of unified tools that guide the agent through a systematic exploration of the codebase, allowing autonomous navigation based on contextual needs. Furthermore, by fine-tuning \texttt{Qwen-2.5-Coder-Instruct}~\cite{hui2024qwen25codertechnicalreport} 7B and 32B models(abbr. as \texttt{Qwen-2.5-7B} and \texttt{Qwen-2.5-32B} respectively), our system achieves performance comparable to state-of-the-art models like \texttt{Claude-3-5-sonnet-20241022}~\cite{anthropic2023claude} (abbr. as \texttt{Claude-3.5}) while significantly reducing API costs by over 80\% (from \$0.66 to \$0.09 per example), making it practical for real-world deployment.

Additionally, to facilitate a comprehensive evaluation of code localization methods, we introduce \textsc{Loc-Bench}, a new benchmark specifically designed for this task. Existing benchmarks like SWE-Bench present significant limitations: (1) they risk contamination through data overlap with LLM training sets \cite{mundler2024swt}, and (2) they primarily focus on bug fixing, lacking diversity in maintenance scenarios such as feature requests, performance optimizations, and security fixes.
In contrast, \textsc{Loc-Bench} covers diverse scenarios and mitigates potential contamination concerns by incorporating more recent examples from popular Python repositories collected after known LLM training cutoff dates.
%offers broader coverage of maintenance scenarios and repository types. 
% We collected 560 examples from popular Python repositories, balanced across bug reports (242), feature requests (150), security issues (29), and performance problems (139), creating a diverse evaluation environment that better reflects real-world code maintenance challenges. 
Additionally, we provide tooling to continuously update the benchmark with new examples, allowing researchers to maintain a fresh evaluation dataset as models evolve and training data cutoffs advance.

Our contributions address critical gaps in existing approaches:

\begin{itemize}
    \item We introduce a heterogeneous graph representation that captures both explicit and implicit code relationships, enabling efficient multi-hop reasoning. Our lightweight graph-based indexing process takes only seconds per repository and requires minimal storage.
    \item We design unified tools for agent-based code exploration that leverage our graph representation, allowing LLM agents to perform complex multi-hop navigation and reasoning across code dependencies even when target code isn't explicitly mentioned in issue descriptions.
    \item We introduce Loc-Bench, a new benchmark specifically designed for code localization that addresses limitations in existing datasets. Unlike previous benchmarks dominated by bug reports, Loc-Bench offers a balanced distribution across bug fixes, feature requests, security patches, and performance optimizations.
    \item By fine-tuning open-source models on this task, we reduce the cost of code localization by 86\% while maintaining competitive performance.

\end{itemize}
% contamination-free 

% Graph Traversal Perspective 
% Question 2: Challenges in exisiting code localization
% Why we want to construct a graph? We want to easily perform multi-hop traversal.
% 1. step-by-step traversal is time consuming, inefficient, and noisy.
% 2. hard to leverage useful a priori knowledge.
% 3. graph skeleton

\begin{figure*}[t]
    \centering
    \includegraphics[width=\textwidth, trim=45 272 207 46, clip]{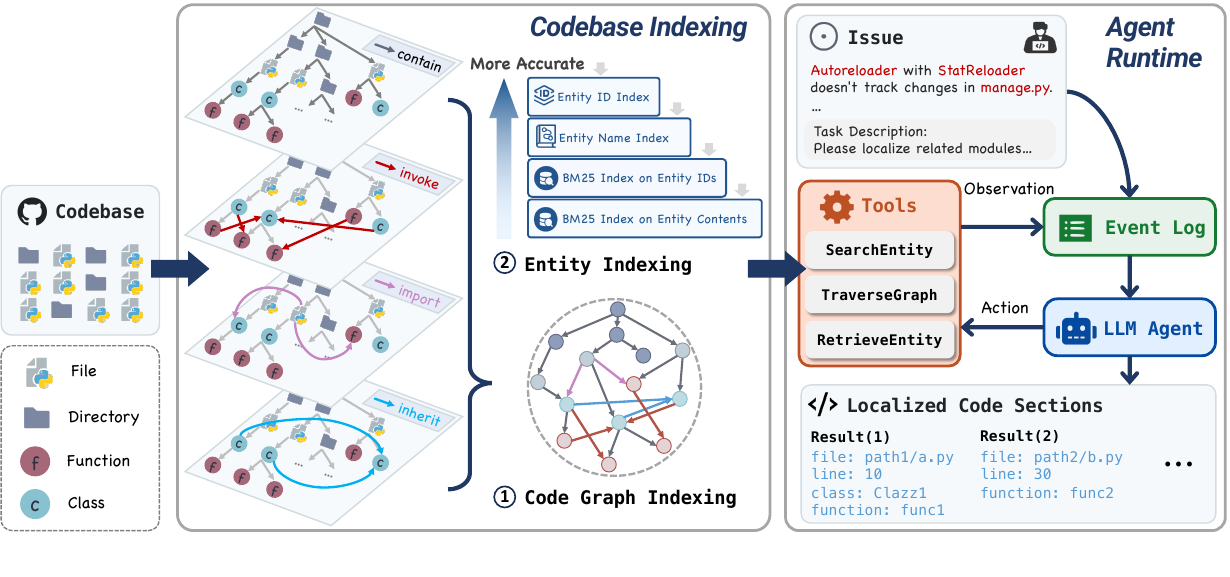} 
    % \vspace{-10mm}
    \caption{Overview of \textsc{LocAgent} framework. \textsc{LocAgent} first parses the given codebase to build a graph-based code representation with various types of entities and relations. It then constructs sparse indexes for exploring structures and searching content. Using these indexes, it performs agent-guided searches that combine the graph and tools.}
    \label{fig:main_fig}
    % \vspace{-.3cm}
    % \hspace{-5mm}
\end{figure*}

\section{Related Work}

%Common solutions rely on traditional IR methods~\cite{wang2023codet5opencodelarge,günther2023jinaembeddingsnovelset,zhang2024coderepresentationlearningscale,suresh2024cornstack}, which are based on similarity matching to return ranked lists of code snippets.
%However, encoding the entire repository is computationally expensive and time-consuming~\cite{de2020autoregressive}, making it difficult to adapt to fast-evolving codebases. Additionally, these methods often fail to capture the fine-grained information of code entities.
% with the advancement of pre-trained language models, generative information retrieval (GenIR) has emerged as a novel paradigm~\cite{de2020autoregressive, tay2022transformer,li2024matching}, greatly reducing memory footprint, but still lacking enough reasoning capability to resolve the bug.
%Recently, LLMs with advanced code reasoning capabilities have demonstrated superior performance by directly processing queries and raw code for fault localization~\cite{kang2023preliminary, wu2023large, xia2024agentless, kang2024quantitative}. However, these methods are restricted in real-world scenarios~\cite{jimenez2023swe} due to limited context windows.
%Some agent-based methods~\cite{yang2024swe,qin2024agentfl,orwall2023moatless} utilize multistep reasoning to enable automated codebase traversal, but highly depend on the directory structure of repositories for navigation, resulting in numerous expensive LLM calls and facing challenges in extracting and understanding complex code structures and dependency relationships from raw code.

\subsection{Traditional Retrieval-based Methods}
% Common solutions rely on 
Traditional IR methods rely on lexical or semantic matching to return ranked lists of code snippets.
Sparse retrievers, such as BM25~\cite{robertson1994okapi,robertson2009probabilistic}, have demonstrated robustness to domain adaptation. Dense retrievers utilize embeddings for improved semantic searching, including models with open checkpoints such as general text embedding models E5-base-v2~\cite{wang2022text} and proprietary APIs~\cite{voyageai2024}.
Code embedding models such as Jina-Code-v2 \cite{günther2023jinaembeddingsnovelset}, Codesage-large-v2 \cite{zhang2024code}, and CodeRankEmbed \cite{suresh2024cornstack}, trained specifically for code related tasks, showing significant performance in Code2Code and NL2Code semantic search tasks.
However, while the embedding models themselves are small, the engineering challenges of maintaining these indexing systems (e.g., storage requirements, update mechanisms, and infrastructure maintenance) make them difficult to adapt to fast-evolving codebases.

\subsection{LLM-based Generative Retrieval Methods}
Recently, LLMs with advanced code reasoning capabilities have demonstrated superior performance by directly processing queries and raw code for code localization~\cite{kang2023preliminary, wu2023large, xia2024agentless, kang2024quantitative}. For example, Agentless \cite{xia2024agentless}, initially designed for automated program repair, uses a simplistic hierarchical localization process powered by LLM.
It employs a straightforward three-phase approach that first localizes relevant code sections before attempting to fix the identified issues, challenging the assumption that complex agent architectures are necessary for effective code understanding and modification tasks.

%However, these methods are restricted in real-world scenarios~\cite{jimenez2023swe} due to limited context windows \xw{why? they are actually doing hierarchical, which is potentially less demanding in context window. Suggest remove the last sentence.}.

Expanding on these techniques, agent-based methods utilize multi-step reasoning to enable automated codebase traversal. 
Specifically, OpenHands~\cite{wang2024openhands} implements a generalist coding agent that supports bash commands like \texttt{grep} and tools for viewing files.
SWE-Agent~\cite{yang2024sweagent} integrates a custom Agent-Computer Interface to support agents to navigate entire repositories. MoatlessTools~\cite{orwall2023moatless} combines an agentic searching loop and semantic search to obtain code locations.
% Notably, few agent-based methods are specifically optimized for code localization tasks. 
%However, agent-based methods highly depend on the directory structure of repositories for navigation and face challenges in extracting and understanding complex code structures and dependency relationships from raw code.
However, existing agent-based methods face two critical limitations: (a) they primarily navigate codebases through directory traversal rather than understanding semantic relationships, (b) and they struggle to extract and reason about complex cross-file dependencies when these relationships aren't explicitly represented in the repository structure. This significantly impairs their ability to locate code that requires modification when the issue involves interactions between structurally distant components in the codebase.

% resulting in numerous expensive LLM calls 
% Agent-based localization methods, including those driven by large language models (LLMs), are tailored for navigating code repositories but require numerous costly LLM calls and lack efficient mechanisms to explore explicit code relationships.  Moreover, they lack efficient mechanisms to explore explicit relationships within the codebase, limiting their overall efficiency.
% by providing the agent with both code search tools and retrieval methods using LLM-constructed queries.
% SWE-agent designs a custom agent-computer interface (ACI) that significantly enhances an agent's ability to create and edit code files, navigate entire repositories, and execute tests and other programs, improving the performance of language model agents to enable automated software engineering. 
% AutoCodeRover further provides the LLM agent with specific code search APIs (e.g., searching methods in a certain class) to iteratively retrieve code context and locate the bug locations. Aider first provides a detailed repository map constructed with static and call graph analysis to the LLM to localize the files that require editing.
% These agents, often used in automated repair workflows, enhance flexibility but require numerous computationally expensive LLM calls.

\subsection{Graph-based Code Representation Methods}
\begin{table*}[ht]
\centering
\resizebox{\textwidth}{!}{
\begin{tabular}{l|cccc|cccc|c}
\toprule
\multirow{2}{*}{Method} & \multicolumn{4}{c|}{Relation Types} & \multicolumn{4}{c|}{Node Types} & \multirow{2}{*}{Search/Traversal Strategy} \\
\cmidrule(lr){2-5} \cmidrule(lr){6-9}
 & Contain & Import & Inherit & Invoke & Directory & File & Class & Function & \\
\midrule \midrule
CodexGraph\cite{liu2024codexgraph} & \textcolor{red}{\halfcheckmark} & \textcolor{red}{\ding{55}} & \textcolor{green}{\ding{52}} & \textcolor{green}{\ding{52}} & \textcolor{red}{\ding{55}}& \textcolor{red}{\ding{55}}& \textcolor{green}{\ding{52}} & \textcolor{green}{\ding{52}} & Cypher queries \\
RepoGraph\cite{ouyang2025repograph} & \textcolor{red}{\halfcheckmark}  & \textcolor{red}{\ding{55}}& \textcolor{green}{\ding{52}} & \textcolor{green}{\ding{52}} & \textcolor{red}{\ding{55}}& \textcolor{red}{\ding{55}}& \textcolor{green}{\ding{52}} & \textcolor{green}{\ding{52}} & Ego-graph retrieval \\
RepoUnderstander\cite{ma2024understand} & \textcolor{green}{\ding{52}} & \textcolor{red}{\ding{55}}& \textcolor{green}{\ding{52}} & \textcolor{green}{\ding{52}} & \textcolor{green}{\ding{52}} & \textcolor{green}{\ding{52}} & \textcolor{green}{\ding{52}} & \textcolor{green}{\ding{52}} & MCTS \\
OrcaLoca\cite{yu2025orcaloca} & \textcolor{green}{\ding{52}} & \textcolor{red}{\ding{55}}& \textcolor{red}{\ding{55}}& \textcolor{red}{\halfcheckmark} & \textcolor{green}{\ding{52}} & \textcolor{green}{\ding{52}} & \textcolor{green}{\ding{52}} & \textcolor{green}{\ding{52}} & Simple search tools \\
\midrule
\textsc{LocAgent}(Ours) & \textcolor{green}{\ding{52}} & \textcolor{green}{\ding{52}} & \textcolor{green}{\ding{52}} & \textcolor{green}{\ding{52}} & \textcolor{green}{\ding{52}} & \textcolor{green}{\ding{52}} & \textcolor{green}{\ding{52}} & \textcolor{green}{\ding{52}} & Unified retrieval tools \\
\bottomrule
\end{tabular}
}
\caption{Comparison of Graph-Based Code Representation Methods.}
\vspace{-.3cm}
\label{tab:graph_repre}
\end{table*}

Due to the inherent structure of code, several works have employed graph-based representations to improve code understanding by capturing key relationships between components.
\citet{aider2025repomap} constructs a RepoMap and uses a graph ranking algorithm to identify the most significant contextual elements.
Similarly, as a plugin, RepoGraph~\cite{ouyang2025repograph} performs subgraph retrieval – extracting an ego-network of relevant lines and their neighbors – to provide structured context.
CodexGraph~\cite{liu2024codexgraph} indexes the repository into a Neo4j graph database, where LLM agents query the database precisely using Cypher. The efficiency of its retrieval process depends heavily on the querying capabilities of the LLM.
These methods focus primarily on providing relevant context but do not enhance the traversal process itself, as they do not explicitly model directory structure or file hierarchies.

In contrast, RepoUnderstander~\cite{ma2024understand} builds hierarchical and function-call graphs, using Monte Carlo Tree Search (MCTS) guided by an LLM for exploration. While thorough, MCTS introduces extra computational overhead, making it less efficient than simpler traversal methods like BFS, particularly in large repositories.
OrcaLoca~\cite{yu2025orcaloca} uses a simplified graph enhanced by priority scheduling and context pruning. It maintains efficient search but may miss complex invocation dependencies.
Table \ref{tab:graph_repre} summarizes the differences between these methods and \textsc{LocAgent}. Compared to these approaches, \textsc{LocAgent} offers a more comprehensive and unified representation of the repository, along with efficient, unified retrieval tools specifically designed for LLM consumption.

\section{The \textsc{LocAgent} Framework}
% 1. introduce how to build graph
% \subsection{Overview}

We introduce \textsc{LocAgent}, a graph-oriented LLM-agent framework for code localization.
Figure~\ref{fig:main_fig} illustrates the overall framework.
When given a repository, \textsc{LocAgent} can locate all the relevant code sections at various granularities (file, class, function, or line level) for different types of GitHub issues (such as bug reports, feature requests, performance bottlenecks, and security vulnerabilities) through automated in-depth exploration and analysis of the codebase.
Section~\ref{sec: graph-interface} proposes a novel graph-based indexing approach as an intermediate representation for codebases.
Section~\ref{sec: agent-search} presents our agent-based code search on the indexes and Section~\ref{sec: fine-tune} describes our model fine-tuning and distillation process.

\subsection{Graph-based Code Representation}\label{sec: graph-interface}
% \subsection{Unified Graph Interface for Agent-Codebase Interaction}\label{sec: graph-interface}

% Outline
% 1. Convert repository to graph: challenges: granularity, edge type, node type
% 2. Provide efficient interaction with graph, 

% provable more expressive, including directory tree (compared with previous methods)
% provable, more efficient graph primitives, graphs include rich information, including interaction entities. LLM analysis on graphs can explore a wide range of code entities while not looking into their implementation. saving time and cost.
% unified, easier to learn, easier to deploy, maintain, and iterate.

% 1. existing work expose repository structure to agent, allow agent to search by keywords, open files, check code structures, by providing numerous tools.
% 2. Tool design: Search for node (entity linking), BFS Traversal on Graph (with Type Filter), Retrieve node content.

% \subsubsection{Graph-based Codebase Indexing}\label{sec: graph-indexing}

Codebases contain rich structural information, both explicit and implicit, that is essential for agent reasoning. 
% For example, files are explicitly organized in a directory-tree, while classes and functions within files can also be implicitly represented as trees. Additionally, the dependency relationships between functions and classes play a crucial role. 
% This motivates us to build a unified graph that comprehensively captures the relationships within a codebase while providing appropriate node granularity for LLM-agents to process.
Building on this insight, we develop a graph-based indexing that comprehensively captures codebase relationships while maintaining a granularity suitable for LLM-agents to retrieve.

\noindent{\textbf{Code Graph Construction}}. We construct a heterogeneous directed graph $\mathcal{G}(\mathcal{V}, \mathcal{E}, \mathcal{A}, \mathcal{R})$ to index the codebase, where $\mathcal{V} = \{ v_i \}^n_{i=1}$ is the node set and $\mathcal{E}\subseteq\mathcal{V}\times \mathcal{V}$ is the edge set.
% Each node \( v \in \mathcal{V} \) and edge \( e \in \mathcal{E} \) are associated with their type mapping functions \( \tau(v):\mathcal{V} \to \mathcal{A} \), \(\mathcal{A}=\{dir, file, class, function\}\) and \( \phi(e):\mathcal{E} \to \mathcal{R}, \mathcal{R}=\{contain, import, invoke, inherit\} \), respectively.
Each node \( v \in \mathcal{V} \) and edge \( e \in \mathcal{E} \) has an associated type mapping function. For nodes, \( \tau(v):\mathcal{V} \to \mathcal{A} \) maps to types \(\mathcal{A}=\{\texttt{directory}, \texttt{file}, \texttt{class}, \texttt{function}\}\). For edges, \( \phi(e):\mathcal{E} \to \mathcal{R} \) maps to relationships \(\mathcal{R}=\{\texttt{contain}, \texttt{import}, \texttt{invoke}, \texttt{inherit}\}\). 
In this paper, we focus our study on Python repositories and leave codebases with other programming languages as future work.

% We first include all directories and Python files as nodes, then parse each Python file using abstract syntax tree (AST) to recursively identify inner functions and classes as nodes. We set the smallest node granularity to the function level and use each function's code content as the \textit{document} to be retrieved by the agent. This approach balances the information density between the index and documents, enabling effective reasoning for LLMs within a limited context window.

First, we include all directories and Python files as nodes. Then, we parse each Python file using the abstract syntax tree (AST) to identify inner functions and classes recursively as nodes. We set the function level as the smallest node granularity and use each function's code content as the document for agent retrieval. This approach creates a good balance of information density between the index and documents, allowing LLMs to reason effectively within their context window limitations.

As shown in Figure~\ref{fig:main_fig}, all nodes with different types can be connected as a single tree using the $contain$ relationship. This structure supports standard codebase-navigation operations from existing works.
% , such as displaying files in a directory, showing classes and functions in a file, and listing member/inner functions within a class/function.
Our code graph further incorporates more advanced codebase relationships as edges:
% Figure~\ref{fig:main_fig} demonstrates three other types of edges: 
(1) the $invoke$ relationship from function/class to function/class, where an invoke to a class represents class instantiation; (2) the $import$ relationship from file to function/class; and (3) the $inherit$ relationship between classes.
% We refer further implementation details to appendix~\ref{sec: append_graph_construct}.

% Our code graph provides a clear and dynamic representation of the codebase architecture, enabling better analysis, understanding, and management of the system's components and their interactions. 
% This allows for type-aware traversal and directional exploration of relationships, such as dependency or inheritance links. These annotations enhance the semantic richness of the traversal process, enabling agents to navigate complex graphs more effectively.

% To mimic how humans analyze the given issue, we also utilize edge indexes to locate the neighbors of the entry nodes.
% unsimilar with the IR-based method, the agent can explore the graph.
% Edge indexing further facilitates efficient traversal by indexing edges based on their key properties. 
% This enables rapid retrieval of all edges associated with a specific relationship type, reducing traversal complexity.

\noindent{\textbf{Sparse Hierarchical Entity Indexing.}}
We treat nodes in our code graph as entities and build hierarchical indexing based on their contents.
% Since issue statements typically include keywords of relevant entities, it's essential to provide an efficient keyword lookup. 
For each keyword, we lookup the indexes from top to bottom:
% If a keyword doesn't match in the upper index, we use the lower index to perform a more fuzzy search.
(1) We build an entity ID index as a unique identifier for each node using its fully qualified name. For example, a function \texttt{calculate\_sum} in the \texttt{MathUtils} class located in \texttt{src/utils.py} would be represented as: \texttt{src/utils.py:MathUtils.calculate\_sum}. 
(2) We construct a global dictionary to map the entity name (e.g., \texttt{calculate\_sum}) to all nodes that share the same name.
(3) We index entity IDs through an inverted index (i.e., BM25) to handle keyword searches that don't exactly match the IDs or names of entities.
(4) For cases where input keywords aren't part of the entities' IDs (e.g., when a keyword refers to a global variable), we build an inverted index that maps code chunk(s) to each entity to cover all possible matches.

% - why graph index designed for agentic search
%     - allowing the LLM agent to navigate relationships within the codebase effectively
% - Sparse Indexing over Semantic Indexing
%     - Sparse Indexing is easier to maintain, keyword matching is better than semantic search. LLM-agents with strong coding capability can better select right entity based on the context.
% - Our graph index is fast to build and easy to update.
\begin{remark}
% Our graph-based indexing unifies code content and structure, enabling efficient multi-hop navigation and keyword search. It's quick to build and update, making it ideal for evolving codebases.
Rather than relying solely on directory structures or hierarchical module indexing, our approach captures module dependencies that transcend directory boundaries.
Two modules in distant directories (A and B) may appear unrelated in traditional navigation, but if they invoke each other or share inheritance, they're syntactically close in our graph representation.
This syntactic proximity is essential for code localization because issues typically manifest through call relationships rather than directory structure. By capturing these functional dependencies, our approach efficiently identifies related components even when physically distant in the codebase.

% Given the strong coding capabilities of LLM agents, they are easy to pick up critical keywords and can better select the appropriate entities based on the context. 
% Additionally, our graph index is lightweight, fast to build, and easy to update, making it an efficient solution for fast-evolving codebases.
\end{remark}

\subsection{Agent-guided Code Search} \label{sec: agent-search}

We develop tools based on the indexes built offline. During runtime, \textsc{LocAgent} takes issue statements as input and launches agents that autonomously use tools to localize target code sections. While the agent may iteratively invoke multiple tools internally to explore the codebase, \textsc{LocAgent} presents a simplified interface to users, requiring only a single-turn interaction—users submit an issue statement and receive localization results without additional input. This autonomous, self-contained workflow makes \textsc{LocAgent} both easy to deploy and highly practical for real-world use.

%\xw{this looks confusing? how can it both be iteratively use trool, then require one-turn interaction?}

% SWE-Agent~\cite{yang2024swe} highlights the importance of an LM-friendly Agent-Computer interface to solve complex software engineering tasks. 
% Recent works~\cite{orwall2023moatless, wang2024openhands}, inspired by GUI-based IDEs, have developed numerous specialized tools for agents to explore codebases. 
% However, these tools are initially designed for human readability, which sacrifices the compactness and efficiency that LLM agents prefer~\cite{yang2024swe}.
% Unlike humans, LLMs excel at processing large volumes of information at once but struggle with higher-order operations with many turns.
% We address these challenges with an innovative graph interface that: (1) structures the codebase into a graph-based index as an intermediate representation and (2) enables efficient codebase exploration through three unified tools.
% This subsection shows how our approach surpasses existing interfaces in both capability and efficiency.

% In the code localization stage, the basic components are a set of tools
% which an LLM agent can use to gather relevant code snippets from the codebase. 

\noindent\textbf{Tool Design for Codebase Exploration.}
Recent works~\cite{orwall2023moatless, wang2024openhands}, inspired by GUI-based IDEs, have developed numerous specialized tools for agents to explore codebases. 
However, these tools are initially designed for human readability, which sacrifices the compactness and efficiency that LLM agents prefer~\cite{yang2024sweagent}.
% Unlike humans, LLMs excel at processing large volumes of information at once but struggle with higher-order operations.
% Thanks to our graph-based code representation, we can develop tools that supports efficient higher-order codebase exploration to address these challenges. 
Building upon our graph-based code representation, we can develop tools that support efficient higher-order codebase exploration to address these challenges. We unify all codebase navigation, search, and view operations into three tools (Table~\ref{tab:tool_list}), introduced as follows.

\begin{table}[t]
\resizebox{\linewidth}{!}{
    \centering
    \begin{tabular}{l l l}
        \toprule
        \textbf{Tool Name} & \textbf{Input Params} & \textbf{Output}\\
        \midrule
        \midrule
            \texttt{SearchEntity} & \textit{Keywords} & \makecell[l]{Related Entities \\ with Code Snippets}\\    
        \midrule
            \multirow{5}{*}{\texttt{TraverseGraph}} & \textit{Start Entity IDs} & \multirow{5}{*}{\makecell[l]{Traversed Subgraph, \\ including Entites \\ and  Relations}} \\
            & \textit{Direction} & \\
            & \textit{Traverse Hops} & \\
            & \textit{Entity Types} & \\
            & \textit{Relation Types} & \\
            % \texttt{start\_entity\_ids}, \texttt{direction}, \texttt{number\_of\_hops}, \\
            % \texttt{entity\_type}, \texttt{relation\_type}    
        \midrule
            \texttt{RetrieveEntity} & \textit{Entity IDs} & \makecell[l]{Complete Code \\ of Specified Entities}\\
        \bottomrule
    \end{tabular}
    }
    \caption{List of unified APIs provided by LocAgent for code search and exploration.}
    \label{tab:tool_list}
    \vspace{-5mm}
\end{table}

\texttt{SearchEntity}:
This tool searches codebases using keywords to locate relevant entities through our Hierarchical Entity Index. When an exact match isn't found in the upper index, the system performs a fuzzy search using the lower index. For each entity found, we return its code snippet in three detail levels: \texttt{fold}, \texttt{preview}, and \texttt{full code} (Figure~\ref{fig:output_format}). This effectively prevents lengthy code context and reduces noise fed into agents.

% issue中有描述这个问题对应的代码位置
% For typical software project issues, we observe that users often mention some “hints” (Fig.~\ref{fig:keyword_entity}) on which part of the codebase is relevant. These hints can be the names of the relevant functions, classes, or files, and sometimes also contain short code snippets. Although these hints may not directly point to the precise location for code modification, they often reveal code context in the project that is relevant to the current issue.
% Tool 输出
% Once invoked by the LLM agent, the retrieval APIs search for files, classes, methods, and code snippets in the code-base, and return the results back to the agent. 

\texttt{TraverseGraph}: This tool performs a type-aware breadth-first search (BFS) on the code graph, starting from input entities and allowing control over both traversal direction and number of hops. This supports agents to perform arbitrary multi-hop codebase navigation through only one action, significantly improving the efficiency compared with existing agent systems.
Note that by allowing agents to select entity types and relation types for each traversal, this tool effectively leverages the LLM agents' coding expertise to generate proper meta paths---a crucial element for heterogeneous graph analysis~\cite{lv2021we}.
For example, by specifying entity types to \{\texttt{class}, \texttt{function}\} and relation types to \{\texttt{contain}, \texttt{inherit}\}, this tool returns the UML diagram.
Additionally, we design an expanded tree-based format for the output subgraph that encodes both relation types and directions  (Figure~\ref{fig:tree_format}).~\cite{fatemi2023talk} demonstrates that LLM performance on graph reasoning depends on the input graph format. Converting a graph into a tree structure encodes topology through the spatial distance between entity names, thereby deriving better performance. 
For detailed comparisons with alternative graph formats, please see Appendix~\ref{sec:study_graph_format}.

\texttt{RetreiveEntity}: This tool retrieves complete entity attributes for each input entity ID, including essential information such as file path, line number, and code content.

\noindent\textbf{Chain-of-Thought Agent Planning.}
% Instead of directly locate code snippets based on the given issue query, we utilize chain-of-thought (CoT) reasoning~\cite{wei2023chainofthoughtpromptingelicitsreasoning} to guide the agent solve this problem step-by-step.
We use chain-of-thought (CoT) prompting (shown in Appendix~\ref{sec:appendix_prompt}) to guide the agent in solving code localization problems step by step.
% and deciding when to end.
The agent systematically follows these steps:
% 1. Extract keywords from issue problem statement.
% - Breakdown Problem Statement into different categories, let agent prioritize the keywords
% - elicit internal knowledge from LLM agents, LLM can be familiar with modules in the code repository so that better identify entity related keywords.
(1) \textit{Keyword extraction.} The agent begins by breaking down the issue statement into different categories
% (Problem description, Error trace, Code to reproduce the bug and Additional context) 
and then extracts relevant keywords that are closely related to the problem.
% 2. Link keywords to code entities.
% - Entities refer to the nodes in the graph (only one). Keywords are usually not complete and may be contained in multiple entities. (e.g., ..)
% It's important to distinguish based on the context. Agents required to search and explore the repository to find more context.
% - formatting entities
(2) \textit{Linking keywords to code entities.}
The agent invokes \texttt{SearchEntity} to complete and clarify each extracted keyword.
% Entities correspond to nodes in the code graph, and 
% Since keywords are often incomplete or ambiguous, the agent needs to search and explore the codebase to gather additional context that helps clarify the issue query.
% the agent proceeds to link them to specific code entities within the repository.
% 3. Reproduce the Problem with an execution flow.
% - identify entry points triggers the issue, trace function calls, identify target entities causing the issue
% - we notice that model will implicity construct execution flow, while achieve high accuracy.
(3) \textit{Generate the logical flow from fault to failure.} 
% After locating the entities, 
The agent first identifies the entry points that trigger the problem. Then, it iteratively traverse the codebase with \texttt{TraverseGraph}, retrieves code contents with \texttt{RetrieveEntity}, and searches new keywords with \texttt{SearchEntity}. Finally, it generates the logic flow based on the issue and additional context.
 % and trace function calls to pinpoint the target entities responsible. 
%, generate the logical flow from fault to failure.
% 4. Locate the target entities.
% - Different granularity, files, functions, lines
% - consider upstream and downstream dependencies
% - output formatting
(4) \textit{Locate the target entities.} The agent pinpoints all suspicious code entities that need modification based on the logic flow. Then, it ranks these entities based on their relevance.
\noindent \textbf{Confidence Estimation Based on Consistency.}
After generating a complete ranked list of candidate entities, to obtain a more consistent ranking, we measure the consistency~\cite{wang2023selfconsistencyimproveschainthought} of the LLM's predictions across multiple iterations. Specifically, we use the Reciprocal Rank as the initial confidence score for each predicted location. We then aggregate the scores for each entity across iterations to compute its final confidence score.
The intuition behind this approach is that if the LLM consistently ranks a location higher in multiple iterations, it is more likely to be relevant.

\subsection{Open-source Model Fine-tuning} \label{sec: fine-tune}
% The model \texttt{claude-3-5}~\cite{anthropic2023claude} demonstrates strong reasoning abilities but is expensive and closed-source. 
Given the high costs of proprietary LLM APIs and data security concerns, we fine-tuned open-source models to improve their code localization capabilities and enable local deployment.
% like \texttt{Qwen}~\cite{hui2024qwen25codertechnicalreport} 
% The base models we use here are \texttt{Qwen2.5-7B} and \texttt{Qwen2.5-32B}.
% using only 433 samples curated with \texttt{claude-3-5-sonnet-20241022}, as detailed in \ref{sec: agent-planning}
% \noindent\textbf{Trajectory Collection.} 
We collect 433 successful trajectories generated with \texttt{Claude-3.5}, where the agent completed tasks from the SWE-bench training set.
%Due to budget constraints, we sample an additional 335 trajectories generated by the fine-tuned \texttt{Qwen2.5-32B} model,% The data from these successful trajectories .these trajectories are used to fine-tune the 32B model further, eventually distilling a smaller 7B model from the entire dataset.
Due to budget constraints, we sample an additional 335 trajectories generated by the initially fine-tuned \texttt{Qwen2.5-32B} model. Importantly, we only select successful trajectories where the model correctly localized the issues, creating a high-quality dataset of correct reasoning paths. These successful examples are then used to refine the same 32B model further, reinforcing effective reasoning patterns through this self-improvement loop. The entire dataset, combining both Claude-3.5 trajectories and successful \texttt{Qwen2.5-32B} samples, was then used to distill knowledge to a smaller 7B model.

% \noindent\textbf{SFT with Lora.}
To fine-tune the smaller model, we employ Supervised Fine-Tuning (SFT) with LoRA~\cite{hu2021lora}.
% and did not include an RL stage; even though incorporating RL could substantially boost model performance, we leave the exploration of the RL stage as feature work. 
Our experiments show that this straightforward distillation method significantly enhances the performance of smaller models.
See Appendix~\ref{appendix_ft_detail} for more training details.

% Each success trajectory, on average, has 12 LM completion messages (roughly 6 turns) and 18, 578 tokens.
% Training on SWE-bench trajectories turns LM into effective agents to fix issues. As shown in Tab.3, despite the base model Qwen-2.5-Coder-Instruct-32B only performs 3\% and 7.0\% on S WE-Bench Lite and Verified， the model fine-tuned on SWE-Gym-sampled trajectories is able to achieve consistent improvements, up to 12.3\% （3.0\% - 15.3\%） and 13.6\% （7.0\% - 20.6\%） absolute performance （i.e.， more Github issue resolved） with the largest 32B model.

% (2 pages including large main figure)
\section{\textsc{Loc-Bench}: A New Benchmark for Code Localization}

\subsection{Revisiting Existing Benchmark}
SWE-Bench\cite{jimenez2023swe} is a widely used benchmark that collects GitHub issues and corresponding code patches that resolve them. \citet{xia2024agentless, suresh2024cornstack} adapt its subset, SWE-Bench-Lite, for code localization, treating the patched files and functions as the targets.
However, existing datasets, including SWE-Bench, present challenges for effectively evaluating code localization methods. First, they are at risk of contamination, as they may include data overlapping with the repositories or issues used by modern models during pre-training.
%Second, existing datasets are not specifically designed for code localization. They primarily focused on bug fixing~\cite{tomassi2019bugswarm}, lack diversity in maintenance scenarios, such as feature requests or performance and security issues, limiting the scope of the evaluation.
Second, existing datasets are not specifically designed for code localization ~\cite{tomassi2019bugswarm}. SWE-Bench, for instance, was created primarily to evaluate end-to-end bug-fixing capabilities, with localization being only an implicit intermediate step. This focus results in datasets dominated by bug reports (85\% of SWE-Bench-Lite examples) while severely under-representing other common software maintenance tasks such as feature requests (14\%), security vulnerabilities (1\%), and performance optimizations (0\%). This imbalance fails to capture the diverse localization challenges faced in real-world software development.
%, where identifying code for feature additions or security hardening requires different reasoning than bug localization.

%\xw{we should in general talk more about this in intro}
% As shown in Table~\ref{tab:dataset_clf}, SWE-Bench\_Lite contains 300 samples, mostly bug reports, and lacks diversity in maintenance scenarios, such as feature requests or performance and security issues.

% Code localization is a static process that takes a repository as a corpus, an issue as a query, and modified files, classes, or functions in the final patch as the ground truth retrieval entities.

% \subsection{Dataset construction Process and Dataset Statistics}
\subsection{Dataset Construction}
% \jialong{missing construction Process details}
To address the limitations of existing benchmarks, we introduce \textsc{Loc-Bench}, a new dataset specifically designed for code localization. This dataset collects up-to-date issues from Python repositories to mitigate the influence of pre-training bias in the latest LLMs.
% Additionally, it includes repositories of varying sizes to ensure robustness across diverse codebases. 
Additionally, \textsc{Loc-Bench} covers wider categories, including bug reports, feature requests, security, and performance issues, enabling a more comprehensive evaluation of code localization methods.
The statistics of \textsc{Loc-Bench} are shown in Table~\ref{tab:dataset_clf}.
% By treating code localization as a static process and using modified files, classes, or functions in the final patch as ground truth, the dataset eliminates the constraints of error reproducibility found in prior benchmarks. 

% \noindent \textbf{Dataset construction} 
% we collected 660 new issues from October 2024 (post-release of Claude-3.5 and GPT-4o) and
For the Bug Report category, we collect GitHub issues created after October 2024, which is later than the release dates of most modern LLMs. To enrich the dataset with more instances of security and performance issues, we use the GitHub Search API to search for relevant keywords, such as "latency improvement" for performance-related issues.
% "Out-of-bounds Write" for security issues and 
We exclude instances that involve modifying more than five Python files or more than ten functions in the corresponding patch. For further details, see Appendix~\ref{appendix_dataset_construction}.

\begin{table}[!t]
\centering
\resizebox{0.95\linewidth}{!}{
\begin{tabular}{l l c}
\toprule
\textbf{Dataset} & \textbf{Category} & \textbf{\#Sample} \\
\midrule
\midrule
\multirow{4}{*}{SWE-Bench-Lite} & Bug Report           & 254 \\ \cmidrule(lr){2-3}
                                 & Feature Request      & 43  \\ \cmidrule{2-3}
                            (Total = 300)     & Security Issue & 3 \\ \cmidrule{2-3}
                                 & Performance Issue    & 0  \\ 
% \midrule

% \multirow{5}{*}{SWE-bench\_Verified} & Bug Report       & 430 \\ \cmidrule{2-3}
%                                  & Feature Request      & 56  \\ \cmidrule{2-3}
%                             (500)     & Security Vulnerability & 2 \\ \cmidrule{2-3}
%                                  & Performance Issue    & 12 \\
\midrule
\multirow{4}{*}{Loc-Bench} & Bug Report       & 242 \\ \cmidrule{2-3}
                                 & Feature Request      & 150  \\ \cmidrule{2-3}
                         (Totoal = 560)        & Security Issue & 29 \\ \cmidrule{2-3}
                                 & Performance Issue    & 139 \\ 
\bottomrule
\end{tabular}
}
\caption{Distribution of samples across different categories in the SWE-Bench-Lite and Loc-Bench datasets.}
\label{tab:dataset_clf}
\vspace{-.4cm}
\end{table}

\begin{table*}[!ht]
    \centering
    \vspace{-.5cm}
    \small
    \resizebox{1\textwidth}{!}{
    \begin{tabular}{l|ll ccc cc cc}
        \toprule
        \multirow{2}{*}{\textbf{Type}} & \multirow{2}{*}{\textbf{Method}} & \multirow{2}{*}{\textbf{Loc-Model}} & \multicolumn{3}{c}{\textbf{File} (\%)} & \multicolumn{2}{c}{\textbf{Module} (\%)} & \multicolumn{2}{c}{\textbf{Function} (\%)}\\
        \cmidrule(lr){4-6} \cmidrule(lr){7-8} \cmidrule(lr){9-10}
         & & & \textbf{Acc@1} & \textbf{Acc@3} & \textbf{Acc@5} & \textbf{Acc@5} & \textbf{Acc@10} & \textbf{Acc@5} & \textbf{Acc@10} \\
        \midrule
        \midrule
        \multirow{5}{*}{Embedding-Based} 
            & \multicolumn{2}{l}{BM25~\cite{robertson1994okapi}} & 38.69    & 51.82 & 61.68  & 45.26    & 52.92    & 31.75	    & 36.86 \\
            & \multicolumn{2}{l}{E5-base-v2~\cite{wang2022text}} & 49.64 & 74.45 & 80.29 & 67.88 & 72.26 & 39.42 & 51.09\\
            % & \multicolumn{2}{l}{CodeT5+~\cite{wang2023codet5opencodelarge}} & 34.67 & 56.93 & 67.15 & 50.73 & 58.39 & 25.55 & 32.85\\
            & \multicolumn{2}{l}{Jina-Code-v2~\cite{günther2023jinaembeddingsnovelset}} & 43.43 & 71.17 & 80.29 & 63.50 & 72.63 & 42.34 & 52.19\\
            & \multicolumn{2}{l}{Codesage-large-v2~\cite{zhang2024code}}  & 47.81 & 69.34 & 78.10 & 60.58 & 69.71 & 33.94 & 44.53\\
            & \multicolumn{2}{l}{CodeRankEmbed~\cite{suresh2024cornstack}} & 52.55 & 77.74 & 84.67 & 71.90 & 78.83 & 51.82 & 58.76 \\
        \midrule
        \multirow{2}{*}{Procedure-Based} 
            & \multirow{2}{*}{\makecell[l]{Agentless\\\cite{xia2024agentless}}}  
                & \texttt{GPT-4o}   & 67.15 & 74.45 & 74.45 & 67.15 & 67.15 & 55.47 & 55.47 \\
            & & \texttt{Claude-3.5} & 72.63 & 79.20 & 79.56 & 68.98 & 68.98 & 58.76 & 58.76 \\
        \midrule
        \multirow{9}{*}{Agent-Based} 
            & \multirow{2}{*}{\makecell[l]{MoatlessTools\\\cite{orwall2023moatless}}} 
                & \texttt{GPT-4o}   & 73.36 & 84.31 & 85.04 & 74.82 & 76.28 & 57.30 & 59.49  \\
            & & \texttt{Claude-3.5} & 72.63 & 85.77 & 86.13 & 76.28 & 76.28 & 64.60 & 64.96  \\
        \cmidrule(lr){2-10}
            & \multirow{2}{*}{\makecell[l]{SWE-agent\\\cite{yang2024sweagent}}}
                & \texttt{GPT-4o}   & 57.30	&64.96	&68.98	&58.03	& 58.03 &	45.99 & 46.35 \\
            &                      & \texttt{Claude-3.5} & 77.37 &	87.23&	90.15&	77.74&	78.10&		64.23&	64.60  \\
        \cmidrule(lr){2-10}
            & \multirow{2}{*}{\makecell[l]{Openhands\\\cite{wang2024openhands}}} 
                & \texttt{GPT-4o}   & 60.95 & 71.90 & 73.72 & 62.41 & 63.87 & 49.64 & 50.36  \\
            & & \cellcolor{gray!10}\texttt{Claude-3.5} & \cellcolor{gray!10}76.28 & \cellcolor{gray!10}89.78 & \cellcolor{gray!10}90.15 & \cellcolor{gray!10}83.21 & \cellcolor{gray!10}83.58 & \cellcolor{gray!10}68.25 & \cellcolor{gray!10}70.07  \\
        \cmidrule(lr){2-10}
            & \multirow{3}{*}{\textsc{LocAgent} (Ours)}
                % & \texttt{GPT-4o}      &63.14	&84.31	&85.77	&74.45	&77.37	&59.85	&65.33  \\
            & \texttt{Qwen2.5-7B(ft)}      & 70.80 & 84.67 & 88.32 & 81.02 & 82.85 & 64.23 & 71.53  \\
            & & \cellcolor{blue!10}\texttt{Qwen2.5-32B(ft)} & \cellcolor{blue!10}75.91 & \cellcolor{blue!10}90.51 & \cellcolor{blue!10}92.70 & \cellcolor{blue!10}85.77 & \cellcolor{blue!10}87.23 & \cellcolor{blue!10}71.90 & \cellcolor{blue!10}77.01  \\
            & & \cellcolor{blue!20}\texttt{Claude-3.5}   & \cellcolor{blue!20}\textbf{77.74} & \cellcolor{blue!20}\textbf{91.97} & \cellcolor{blue!20}\textbf{94.16} & \cellcolor{blue!20}\textbf{86.50} & \cellcolor{blue!20}\textbf{87.59} & \cellcolor{blue!20}\textbf{73.36} & \cellcolor{blue!20}\textbf{77.37} \\
        \bottomrule
    \end{tabular}
    }
    \caption{Performance comparison with baseline methods on code localization on SWE-bench lite. Results show the accuracy at file, module, and function levels. For Agent-Based methods, we use \texttt{GPT-4o-2024-0513} (abbr. as \texttt{GPT-4o}) and \texttt{Claude-3-5-sonnet-20241022} (abbr. as \texttt{Claude-3.5}) as the localization model. Additionally, the performance of our fine-tuned open-source models, \texttt{Qwen2.5-7B(ft)} and \texttt{Qwen2.5-32B(ft)}, are included for comparison.}
    \vspace{-.5cm}
    \label{tab:acc_comparison}
\end{table*}

\section{Experiments}
\vspace{-.1cm}
Our experiments aim to evaluate four key aspects of \textsc{LocAgent}: 
(1) the effectiveness of our graph-based representation and tooling for code localization compared to existing methods, (2) the performance of fine-tuned open-source models as cost-effective alternatives to proprietary LLMs, (3) a detailed analysis of how performance varies across task categories, 
and (4) the contribution of each component in our framework through comprehensive ablation studies. 
We evaluate on both SWE-Bench-Lite and our introduced Loc-Bench dataset. Additionally, we examine the impact of improved localization on downstream software maintenance tasks.

\vspace{-.2cm}

\subsection{Experimental Settings}
\textbf{Datasets.}
We first conduct experiments on SWE-Bench-Lite, treating the patched files and functions as the targets for localization.
Following \citet{suresh2024cornstack}, we excluded examples where no existing functions were modified by the patch, ultimately retaining 274 out of the original 300 examples.
% The statistics of dataset is shown in Appendix~\ref{appendix_dataset_statistics}.

\noindent \textbf{Metrics.}
To assess performance, we use a modified accuracy metric inspired by R-Precision from information retrieval, following Agentless\cite{xia2024agentless}.
To assess performance, we use Acc@k (Accuracy at k) as our evaluation metric, following Agentless\cite{xia2024agentless}. For each example, we select the top-k predicted locations and consider a localization attempt successful only if all relevant locations are correctly identified within these top-k predictions. This approach measures the ability to fully identify all necessary code sections that require modification. We report results across multiple $k$ values: file localization at Acc@1, Acc@3, and Acc@5, and function localization at Acc@5 and Acc@10. Additionally, to provide a more relaxed evaluation criteria, we assess module localization, which only requires finding any function within the patched class.
% This approach adapts to the varying number of required modifications across different issues, providing a fairer evaluation than fixed $k$ cutoffs. 
%For each example, we select the top $k$ predicted locations and consider a localization attempt successful only if \textit{all relevant locations are correctly identified}. 
%This metric best captures the ability to fully identify relevant locations, making it ideal for real-world applications. 
%To evaluate this task fully, we assess file localization at top-1, top-3, and top-5 ranks and function localization at top-5 and top-10. To provide a more relaxed metric for function localization, we also evaluate module localization, which only requires finding any functions among the patched class.

\subsection{Baselines}
We evaluate \textsc{LocAgent} against three categories of competitive baselines:
(a) Retrieval-based methods: We include the sparse retrieval approach BM25 \cite{robertson1994okapi} and several state-of-the-art embedding models, including the general-purpose E5-base-v2 \cite{wang2022text} and specialized code embedding models such as Jina-Code-v2 \cite{günther2023jinaembeddingsnovelset}, Codesage-large-v2 \cite{zhang2024code}, and the current SOTA code embedding model CodeRankEmbed \cite{suresh2024cornstack}. Proprietary embedding solutions were excluded due to API costs.
(b) Procedure-based methods: We compare against Agentless \cite{xia2024agentless}, which employs a structured hierarchical approach to code localization without complex agent architectures.
(c) Agent-based methods: We include several advanced agent frameworks designed for code exploration and modification, specifically OpenHands \cite{wang2024openhands} (using its default CodeActAgent implementation), SWE-Agent \cite{yang2024sweagent}, and MoatlessTools \cite{orwall2023moatless}.
\noindent  For implementation details, please refer to Appendix~\ref{appendix_exp_impl}.
% OpenHands\cite{wang2024openhands}(we use the default CodeActAgent), a generalist coding agent that supports bash commands like \texttt{grep} and tools for viewing files. We also evaluate SWE-Agent, which integrates a carefully designed Agent-Computer Interface. Additionally, we examine MoatlessTools \cite{orwall2023moatless}, which leverages an agentic searching loop functioning as a finite state machine and incorporates semantic search to identify relevant files or functions based on model queries.

\begin{figure}[t]
    \centering
     \vspace{-3mm}
    \includegraphics[width=\columnwidth]{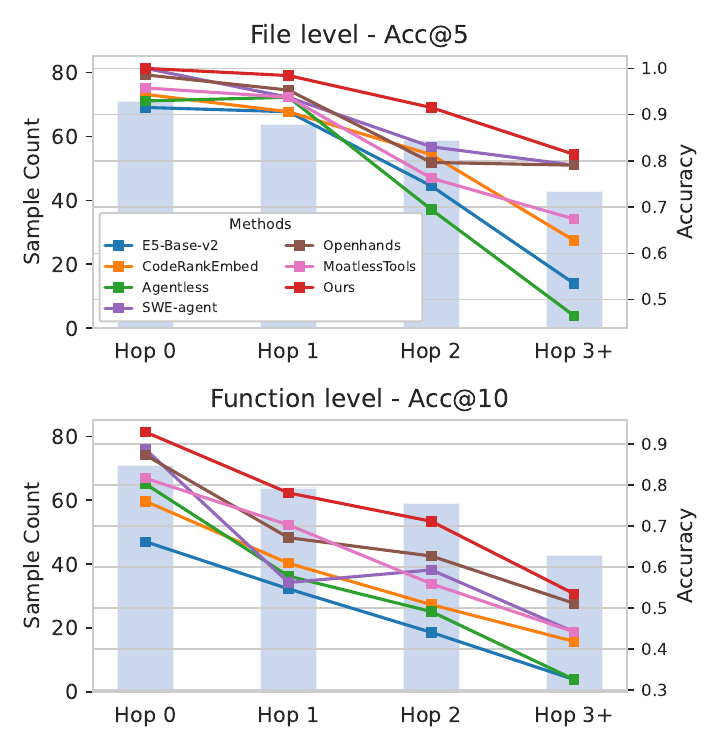}\vspace{-.2cm}
    \caption{Performance analysis at different difficulty levels for file- and function-level localization. All agent-based methods and Agentless use \texttt{Claude-3.5} as the localization model. \textit{Hop $N$} refers to the distances between functions mentioned in the issue description and the ground truth patch on our code graph.
    % \xw{we should add error bar for each of these point - otherwise we don't know the sample count for each hop}
    }
    \vspace{-.2cm}
    \label{fig:acc_among_hops}
    % \hspace{-3mm}
\end{figure}
\subsection{Evaluation Results on SWE-Bench-Lite}
As shown in Table~\ref{tab:acc_comparison}, Agent-Based methods consistently outperform other approaches, and our method demonstrates competitive performance by achieving the best results across all levels of code localization.
Unlike traditional retrieval-based methods, Agentless identifies only a limited number of locations due to its narrow repository scope, which hinders performance gains when considering a broader set of candidates.
The results of the NDCG are presented in Table~\ref{tab:ndcg_comparison} in the Appendix.

% Openhands shows strong performance with \texttt{Claude-3.5}, benefiting from a well-designed tool specifically tailored for this model. However, when switched to \texttt{GPT-4o}, its performance drops significantly across all metrics comparing to ours, especially in module and function localization. This suggests that Openhands heavily relies on the compatibility between its tool and the LLM in use. Similarly, MoatlessTools exhibits relatively stable performance with \texttt{GPT-4o}, but its improvement when switching to the more advanced \texttt{Claude-3.5} is marginal, indicating limited adaptability of its tool to stronger models. 
% In contrast, our method achieves robust results with both \texttt{GPT-4o} and \texttt{Claude-3.5}, demonstrating consistent adaptability across different LLMs.

To further analyze the results, we examine performance across different task difficulty levels. We measure the task difficulty by calculating the shortest hops between the functions mentioned in the issue descriptions and the patched functions on our code graph (See Appendix~\ref{appendix_diff_dist} for more details). As shown in Figure~\ref{fig:acc_among_hops}, performance decreases for all methods as the task becomes more challenging.
However, Agent-based methods demonstrate better robustness as the difficulty increases, with our method maintaining competitive performance across various difficulty levels.
% All methods achieve over 90\% accuracy in file-level localization when the ground truth functions are explicitly mentioned in the query (\textit{Hop $0$} in the figure). 
Retrieval-based methods, such as E5-Base-v2 and CodeRankEmbed, perform poorly at the function level, even when the patched functions are explicitly mentioned in the query. This is because they treat the query as a whole, failing to capture fine-grained details.
Agentless performs even worse than retrieval-based methods when exploration beyond the query is needed (\textit{hop} $\geq 0$) due to its simplistic localization process and limited view focused only on the repository structure. 
%\xw{I think we should really highlight the advantage of our approach more in abs and intro: the ability to perform multi-hop reasoning like this, whereas prev approaches failed}

\subsection{Fine-tuned Open-source Models}\label{sec:exp_finetune}
\begin{figure}[!tb]
    \centering
    \vspace{-.4cm}   
    \includegraphics[width=\columnwidth]{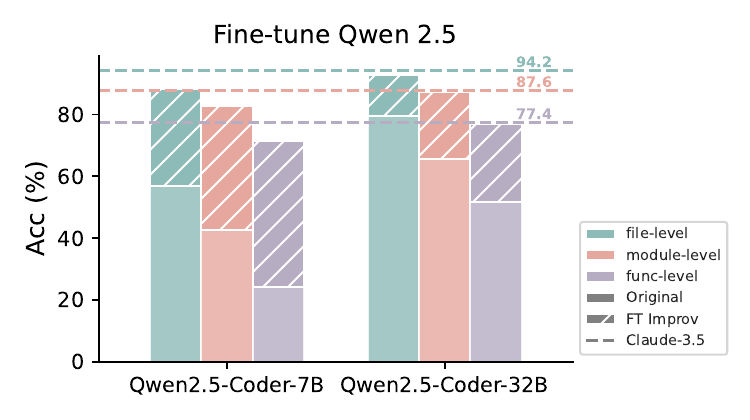}
    \caption{Comparison of performance between the original and fine-tuned Qwen models. The metrics used are file-level Acc@5 and module/function-level Acc@10. Dashed lines represent the performance of the \texttt{Claude-3.5} model for reference.}
    \label{fig:ft_qwen}
    \vspace{-.4cm}
\end{figure}

Figure~\ref{fig:ft_qwen} demonstrates that after fine-tuning, both the 7B and 32B models show significant improvements on this task. \textsc{LocAgent} with fine-tuned \texttt{Qwen-2.5-Coder-Instruct-32B} (abbreviated as \texttt{Qwen2.5-32B(ft)}) achieves performance comparable to \texttt{Claude-3.5}, and \textsc{LocAgent} with \texttt{Qwen2.5-7B(ft)} also delivers results on par with that obtained using \texttt{GPT-4o}.
As shown in Table~\ref{tab:acc_comparison}, our method with \texttt{Qwen2.5-32B(ft)} outperforms nearly all baselines, including those that use larger and more powerful LLMs. The original 7B model performs poorly due to its limited tool-use capability \cite{chen2024teval}. These results validate the feasibility of deploying our fine-tuned open-source models as promising alternatives to proprietary APIs, especially in resource-constrained applications.

% results on swe-bench

\begin{table}[ht!]
    \centering
    \small\hspace{-3mm}
    \resizebox{\linewidth}{!}{
    \begin{tabular}{l l ccc}
        \toprule
        \textbf{Method} & \textbf{LM} & \textbf{\#Round} & \textbf{Cost(\$)} & $\displaystyle \frac{\textbf{Acc@10}}{\textbf{Cost}}$ \\
        \midrule
        \midrule
        % Agentless & gpt-4o & 3 & 0.09 & 617\% \\
        % Agentless & claude-3.5 & 3 & 0.10 & 600\% \\
        \multirow{2}{*}{MoatlessTools} & \texttt{GPT-4o} & 5 & 0.46 & 1.3 \\
         & \texttt{Claude-3.5} & 5 & 0.46 & 1.4 \\
        \midrule
        \multirow{2}{*}{SWE-agent} & \texttt{GPT-4o} & 8 & 0.56 & 0.8 \\
         & \texttt{Claude-3.5} & 9 & 0.67 & 1.0 \\
        \midrule
        \multirow{2}{*}{Openhands} & \texttt{GPT-4o} & 15 & 0.83 & 0.6 \\
         & \texttt{Claude-3.5} & 13 & 0.79 & 0.9 \\
        \midrule
        % Ours & gpt-4o & 9 & 1.09 & 59\% \\
        \multirow{3}{*}{Ours} 
         % & \texttt{GPT-4o} & 9 & 0.91 & 0.7 \\
         & \texttt{Claude-3.5} & 7 & 0.66 & 1.2 \\
         &\texttt{Qwen2.5-7B(ft)} & 6 & 0.05 & \textbf{13.2} \\
         &\texttt{Qwen2.5-32B(ft)} & 9 & 0.09 & 8.6 \\
        \bottomrule
    \end{tabular}
}\vspace{-.2cm}
    \caption{Efficiency analysis comparing the average cost and number of agent interaction rounds required by different methods. The cost-efficiency of each method is evaluated using the ratio of function-level Acc@10 to average cost.}    \vspace{-.3cm}

    % \gd{Remove token, add Acc/Cost}
    \label{tab:cost_efficiency}
\end{table}

\begin{table}[t]
    \centering
    \small
    \resizebox{0.48\textwidth}{!}{%
    \begin{tabular}{lc c c}
        \toprule
        \textbf{Model Setting} & \makecell[c]{ \textbf{File} \\
            \textbf{Acc@5}}  & \makecell[c]{\textbf{Module} \\
            \textbf{Acc@10}} & \makecell[c]{\textbf{Function} \\
            \textbf{Acc@10}} \\
        \midrule
        \midrule
        % Ours & 84.67 & 81.02 & 64.23 \\ % 3/ 5 / 5
        Ours & 88.32 & 82.85 & 71.53 \\
        \midrule
        % w/o \texttt{TraverseGraph} & 83.94 & 76.64 & 61.68   \\ % 3/ 5 / 5
        % Fixed \textit{Relation}=["\textit{contain}"] & 72.99 & 67.52 & 52.55 \\
        % Fixed \textit{Hops}=1 & 83.58 & 77.01	 & 60.22 \\
        w/o \texttt{TraverseGraph} & 86.13 & 78.47 & 66.06   \\
        \rowcolor{gray!10} \textit{Relation Types}: \textit{contain} & 86.50 & 79.56 & 66.42 \\
        \rowcolor{gray!10} \textit{Traverse Hops}: 1 & 86.86 & 80.29	 & 66.79 \\
        
        % graph output format & - & - & - \\
        \midrule
        % w/o \texttt{RetrieveEntity} & 83.91 & 79.20 & 62.04 \\
        w/o \texttt{RetrieveEntity} & 87.59 & 81.39 & 69.34 \\
        
        \midrule
        % w/o \texttt{SearchEntity} & 67.52 & 59.85 & 47.08 \\
        % \rowcolor{red!10}
        w/o \texttt{SearchEntity} & 68.98 & 61.31 & 53.28 \\
        \rowcolor{gray!10} w/o BM25 index & 75.18 & 68.98 & 60.22 \\
        % \rowcolor{gray!10} w/o entity id index &  &  & \\
        \bottomrule
    \end{tabular}
    }
    \caption{The ablation study of our model. The metrics used here are file-level Acc@5, module-level Acc@10, and function-level Acc@10. The impact of removing or fixing components is analyzed to observe how each component contributes to the overall accuracy.}
    \label{tab:ablation_study}    \vspace{-.3cm}

\end{table}

% \begin{table}[ht!]
%     \centering
%     \small
%     \resizebox{0.48\textwidth}{!}{%
%     \begin{tabular}{lcc cc cc}
%         \toprule
%         Method & \multicolumn{2}{c}{File} & \multicolumn{2}{c}{Module} & \multicolumn{2}{c}{Function} \\
%         \cmidrule(lr){2-3} \cmidrule(lr){4-5} \cmidrule(lr){6-7}
%         % & Acc@1 & Acc@3 & Acc@5  & Acc@3 & Acc@5 & Acc@10 & Acc@3 & Acc@5 & Acc@10 \\
%         & Acc@1 & Acc@3 & Acc@3 & Acc@5 & Acc@3 & Acc@5 \\
%         \midrule
%         Ours & 70.3\% & 84.3\% & 72.3\% & 76.3\% & 77.3\% & 61.7\%\\
%         \midrule
%         w/o structure tool & - & - & - & - & - & -  \\
%         structure\_tool(hop=1) & - & - & - & - & - & -  \\
%         structure\_tool(edge=contain) & - & - & - & - & - & -  \\
%         graph output format & - & - & - & - & - & -  \\
%         \midrule
%         w/o search\_entity\_tool & - & - & - & - & - & -  \\
%         \midrule
%         w/o search\_tool & - & - & - & - & - & -  \\
%         w/o bm25 index & - & - & - & - & - & -  \\
%         \bottomrule
%     \end{tabular}
%     }
%     \caption{Comparison of Localization Methods Across Different Metrics (Accuracy)}
%     \label{tab:acc_comparison}
% \end{table}
\subsection{Efficiency Analysis}
% - cost/turns/token?/time effciency analysis
% The agentless method demonstrates the lowest cost and token usage, with only 6 conversation rounds for both GPT-4o and Claude-3.5. However, this simplicity results in limited performance due to the lack of advanced retrieval mechanisms. 
Table~\ref{tab:cost_efficiency} presents an efficiency analysis comparing agent-based methods in terms of cost and the number of agent interactions required. 
MoatlessTools demonstrates good cost-efficiency and requires relatively fewer rounds of interaction. However, the dense embeddings it uses make it difficult and slow to adapt to fast-evolving codebases.
SWE-agent and Openhands also show moderate costs but still do not match the efficiency of \textsc{LocAgent}. For \textsc{LocAgent} with \texttt{Claude-3.5}, although more rounds of interaction are required, the cost remains lower than that of Openhands, illustrating the token efficiency of our tool's outputs.
\textsc{LocAgent} with fine-tuned Qwen models stands out for its superior efficiency\footnote{We calculate the cost based on the prices from AI inference providers \cite{hyperbolic2025,artificialanalysis2025}. Specifically, for the \texttt{Qwen2.5-32B(ft)} model, the cost is \$0.20/1M tokens for both input and output. For the \texttt{Qwen2.5-7B(ft)} model, the cost is \$0.14/1M tokens for input and \$0.28/1M tokens for output.}.
\texttt{Qwen2.5-7B(ft)} is the most cost-efficient option, requiring only \$0.05 per example, while \texttt{Qwen2.5-32B(ft)} offers a more cost-effective alternative to \texttt{Claude-3.5}. These results highlight the potential of fine-tuned open-source models as efficient alternatives, providing an optimal balance of cost-effectiveness and performance that surpasses other methods.

\subsection{Ablation Study}
We conduct an ablation study to evaluate the effectiveness of each component of our toolsets. Due to budget constraints, we use the fine-tuned \texttt{Qwen-2.5-7B} as the localization model for these experiments.

\textit{(1) Each tool in our toolset plays a critical role in code localization performance.} As shown in Table~\ref{tab:ablation_study}, removing any tool, especially the \texttt{SearchEntity} tool, leads to varying degrees of accuracy degradation, particularly in module and function level localization. This highlights the critical role each tool plays in identifying relevant modules and functions.

\textit{(2) The graph structure provides essential information for accurate code localization.} Removing \texttt{TraverseGraph} tool decreases module and function level performance since the agent cannot obtain any structure information about the codebase and relies on reasoning capability to identify call relationship or directory structure. Adding \textit{contain} relationship provides only marginal improvements compared to fully removing \texttt{TraverseGraph}, emphasizing the importance of the other three relationship types and explaining why our method surpasses others relying only on the repository structure.

\textit{(3) Multi-hop exploration is crucial for deep code understanding.} When compared to the full setting, fixing \textit{Hops=1} leads to a moderate decline in file and module-level accuracy, but it causes a more significant decrease in function-level accuracy, underscoring the importance of multi-hop exploration for identifying relevant entities.

\textit{(4) Sparse indexing significantly enhances localization performance.} Removing \texttt{SearchEntity} tool, or even partial removal of its index, causes a substantial drop in performance across all metrics. This demonstrates the effectiveness of building a sparse index on our code graph for improving localization performance.

\subsection{Evaluation Results on Loc-Bench}
\begin{table*}[htbp]
    \centering
    \resizebox{0.92\textwidth}{!}{
    \begin{tabular}{@{}lc ccc cc cc@{}}
        \toprule
        \multirow{2}{*}{Method} & \multirow{2}{*}{Loc Model} & \multicolumn{2}{c}{\textbf{File} (\%)} & \multicolumn{2}{c}{\textbf{Module} (\%)} & \multicolumn{2}{c}{\textbf{Function} (\%)} \\
        \cmidrule(lr){3-4} \cmidrule(lr){5-6} \cmidrule(lr){7-8}
        & & Acc@5 & Acc@10 & Acc@10 & Acc@15 & Acc@10 & Acc@15 \\
        \midrule \midrule
        IR-Based & CodeRankEmbed & 74.29&	80.89&	63.21&	67.50&	43.39&	46.61 \\ 
        Agentless & \texttt{Claude-3.5} & 67.50	& 67.50	& 53.39	&53.39	&42.68	& 42.68 \\ 
        OpenHands & \texttt{Claude-3.5} & 79.82	& 80.00	&68.93	&69.11	&59.11	&59.29 \\ 
        SWE-agent & \texttt{Claude-3.5} & 77.68	& 77.68 & 63.57	&63.75	&51.96	&51.96 \\ 
        \midrule
        \multirow{2}{*}{LocAgent (Ours)}
            & \texttt{Qwen2.5-7B(ft)} & 78.57	&79.64	&63.04	&63.04	&51.43	&51.79 \\
            % &\texttt{Qwen2.5-32B(ft)} &  &&&&&& \\
            &\texttt{Claude-3.5} & \cellcolor{blue!20}\textbf{83.39}	&\cellcolor{blue!20}\textbf{86.07}	&\cellcolor{blue!20}\textbf{70.89}	&\cellcolor{blue!20}\textbf{71.07}	&\cellcolor{blue!20}\textbf{59.29}	&\cellcolor{blue!20}\textbf{60.71} \\
        
        \bottomrule
    \end{tabular}
    }
    \caption{Performance evaluation on the real-world LocBench dataset.}
    \label{tab:loc_res_new_bench}
\end{table*}
To ensure the robustness and generalization of our methods and fine-tuned Qwen models, and to eliminate potential data leakage, we evaluate our new dataset. Since Loc-Bench includes examples that edit 1 to 5 files, we assess file localization at top-5 and top-10 ranks, and function/module localization at top-10 and top-15 ranks.
Table~\ref{tab:loc_res_new_bench} shows that our fine-tuned \texttt{Qwen2.5-7B} model exhibits strong generalization capabilities, maintaining competitive performance compared to SWE-agent using more expensive and strong model.
% For Top-N File = 1, the fine-tuned Qwen2.5 model achieves similar results across ranking methods, with file-level accuracy ranging from 67.70\% to 68.60\%. Though slightly lower than Claude-3.5, it performs consistently well at the module and function levels. For Top-N File = 3 and Top-N File = 5, the fine-tuned Qwen2.5-7B model delivers strong performance, achieving file-level recalls of 81.60\%–82.10\% and 86.20\%–86.70\%, respectively, closely matching the trends of Claude-3.5 while maintaining lower module and function-level performance.
These results highlight the practicality of the fine-tuned \texttt{Qwen2.5-7B} model for real-world applications. Despite being an open-source alternative, it achieves a performance comparable to \texttt{Claude-3.5}, supporting its feasibility as a cost-effective substitute for commercial models in practical scenarios. 

\begin{figure}[!tb]
    \centering
    \vspace{-3mm}
    \includegraphics[width=\columnwidth]{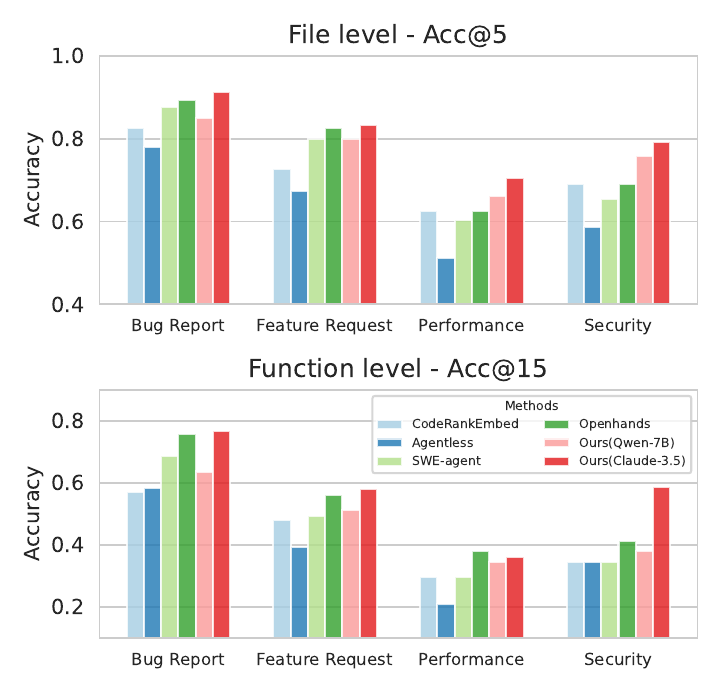}
    \vspace{-7mm}
    \caption{Performance analysis at different difficulty category for file- and function-level localization. All agent-based baselines and Agentless use \texttt{Claude-3.5} as the localization model.}
    \label{fig:acc_on_categories}   
    \vspace{-.3cm}

\end{figure}
Additionally, we evaluate the performance across four different difficulty categories. Figure~\ref{fig:acc_on_categories} clearly shows that our method outperforms other methods in almost all categories of code localization. However, it also highlights a noticeable decrease in performance across the other three categories compared to the Bug Report category. This performance gap likely reflects our training data distribution, which contained more bug report examples, potentially leading to scaffolds better optimized for bug localization tasks. This trend suggests that while our method is highly effective for bug report localization, there is still room for improvement in handling the other categories through more balanced training data and category-specific optimization strategies. 
% For results on full Loc-Bench, see Table~\ref{tab:loc_res_new_bench}.

% loc + agentless
\begin{table}[!tb]
\centering
\resizebox{\linewidth}{!}{%
\begin{tabular}{@{}l l c c c@{}}
\toprule
\textbf{Method} & \textbf{Localization LM} & \textbf{Acc@5} & \textbf{Pass@1} & \textbf{Pass@10} \\ 
\midrule
\midrule
% \multirow{2}{*}{Agentless}  
Agentless
    % & gpt-4o    & 52.9 & 19.07 & 30.33  \\
    & \texttt{Claude-3.5}     & 58.39 & 26.31 & 33.58  \\ 
  \midrule
\multirow{2}{*}{Ours}  
    % & gpt-4o     & 54.7 & - & \\
  % &Qwen2.5-7B (ft)   &  &    &    \\
  &\texttt{Qwen2.5-32B(ft)}    & 69.34 &  26.79  &  36.13  \\ 
  &\texttt{Claude-3.5}   & \textbf{73.36}  & \textbf{27.92} & \textbf{37.59} \\ 
  \bottomrule
\end{tabular}
}
\caption{Impact of localization accuracy on downstream bug repair tasks.}
\label{tab:downstream_edit}    \vspace{-.3cm}

\end{table}

\subsection{Application: Better Localization Leads to More Solved GitHub Issues}
% an automated approach to solving software development problems that ranks among the top-performing open-source submissions on SWE-Bench-Lite. 
To assess the impact of localization methods on downstream tasks, we evaluated their effectiveness in solving GitHub issues.
We choose Agentless as the baseline, ranking among the top-performing open-source submissions on SWE-Bench-Lite.
For consistency, we utilized \texttt{Claude-3.5} as the editing model in conjunction with the Agentless editing method. Table~\ref{tab:downstream_edit} shows that the success rate for solving GitHub issues improves significantly with better code localization accuracy.
% (3 pages)

\section{Conclusion}
In conclusion, \textsc{LocAgent} enhances code localization by structuring codebases as graphs, enabling efficient repository-level exploration for LLM agents. With fine-tuned open-source models, our method achieves high localization accuracy while significantly reducing costs compared to larger proprietary models. Experimental results demonstrate the effectiveness of \textsc{LocAgent} in identifying relevant code components and improving downstream tasks. 
% , making it a cost-effective solution for real-world software development.

\section*{Limitations}

First, our study primarily focused on fine-tuning \texttt{Qwen-2.5-Coder} models. Exploring a broader range of base models, including other open-source LLMs like CodeLlama, Mistral, or Yi, could provide valuable insights into model selection trade-offs. Additionally, investigating different fine-tuning approaches beyond LoRA, such as full fine-tuning or other parameter-efficient methods, could potentially yield better performance.

Second, though we demonstrated improved bug repair performance with better localization, we only scratched the surface of potential downstream applications. Future work should evaluate LocAgent's impact on other software engineering tasks like refactoring, feature addition, security vulnerability patching, and performance optimization. This would provide a more comprehensive understanding of the framework's practical utility.

Moreover, our fine-tuning process relied heavily on trajectories generated by \texttt{Claude-3.5} and the fine-tuned \texttt{Qwen2.5-32B} model. A more diverse training dataset incorporating examples from different models, tasks, and repositories could improve the robustness and generalization of fine-tuned models. Additionally, analyzing the impact of different dataset compositions and filtering strategies on model performance could yield valuable insights.

Finally, the current evaluation focuses primarily on Python codebases. Extending \textsc{LocAgent} to support other programming languages and evaluating its performance across different language paradigms would better demonstrate its generalizability. Further, our evaluation metrics could be expanded to include more nuanced measures of localization quality beyond accuracy and NDCG.

% \section*{Acknowledgments}

% Bibliography entries for the entire Anthology, followed by custom entries
%\bibliography{anthology,custom}
% Custom bibliography entries only
\bibliography{custom}

\newpage
\appendix
\clearpage
\section{\textsc{LocAgent} Design Details}
\begin{figure*}[htb!]  % 使用 figure* 环境
    \centering
    \includegraphics[width=0.98\textwidth, trim=46 235 145 40, clip]{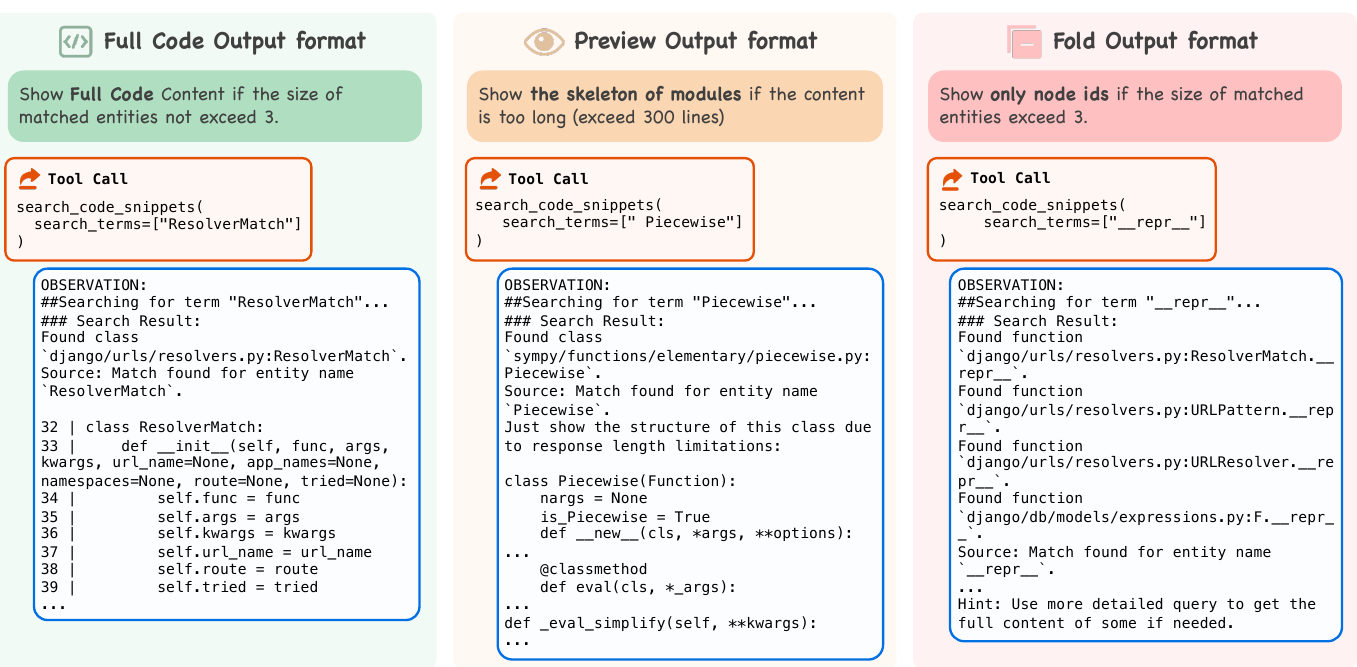}  % 使用 \textwidth 占满整个宽度
    \caption{Different output formats designed for efficient agent-code interaction. Left: Full code output when matched entities $\leq$ 3. Middle: Preview output showing module skeleton for large files. Right: Fold output showing only entity IDs when matches > 3.}
    \label{fig:output_format}
\end{figure*}

\begin{figure*}[htb!]  % 使用 figure* 环境
    \centering
    \includegraphics[width=0.6\textwidth]{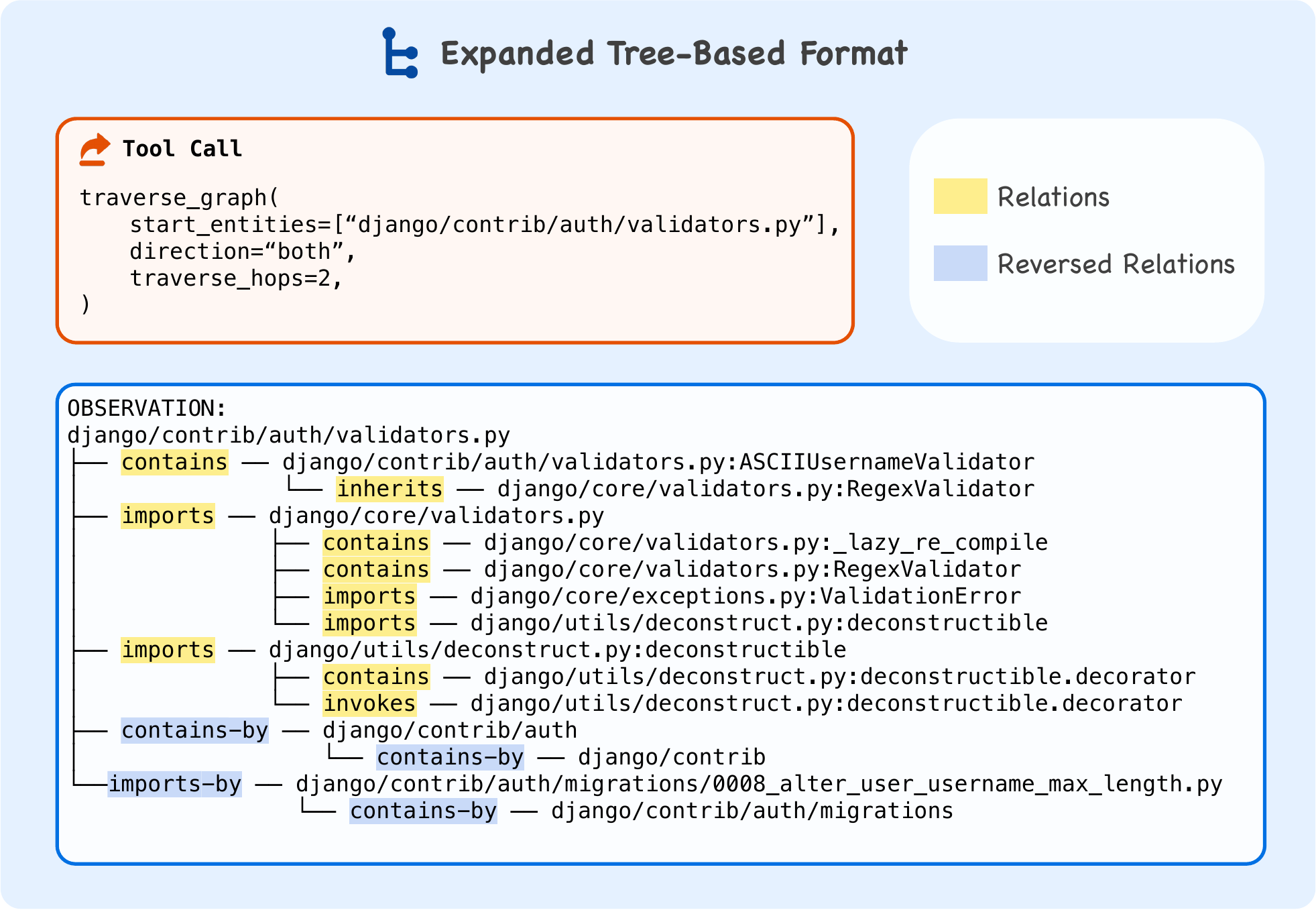}
    \caption{A truncated example of the expanded tree-based format for the output subgraph of tool \texttt{TraverseGraph}.}
    \label{fig:tree_format}
\end{figure*}
\begin{table*}[htbp]
\centering
\resizebox{0.9\linewidth}{!}{
\begin{tabular}{lccc cc cc}
\toprule
\multirow{2}{*}{\textbf{Output Format}} & \multicolumn{3}{c}{\textbf{File(\%)}} & \multicolumn{2}{c}{\textbf{Module(\%)}} & \multicolumn{2}{c}{\textbf{Function(\%)}} \\
\cmidrule(lr){2-4} \cmidrule(lr){5-6} \cmidrule(lr){7-8}
 & Acc@1 & Acc@3 & Acc@5 & Acc@5 & Acc@10 & Acc@5 & Acc@10 \\ \midrule \midrule
row & 41.18 & 67.65 & 70.59 & 61.76 & 61.76 & 35.29 & 38.24 \\ 
row (w/ entity attributes) &41.18 & 64.71 & 64.71& 50.00 & 50.00 & 32.35 & 32.35 \\
incident & 41.18 & 70.59 & 73.53 &55.88 & 55.88 & 29.41 & 32.35 \\ 
% auto\_graph\_struct\_raw &38.24 & 61.76 & 64.71 & 52.94 & 52.94 & 41.18 & 44.12 \\ \midrule
Graphviz DOT &41.18 & 73.53 & \textbf{82.35} & 64.71 & 64.71 & 35.29 & 35.29 \\ 
JSON &41.18 & 67.65 & 76.47 & \textbf{67.65} & \textbf{70.59} & \textbf{38.24} & \textbf{41.18} \\
\textbf{tree-based (Ours)} &\textbf{47.06} & \textbf{79.41} & 79.41 & 64.71 & 64.71 & \textbf{38.24} & \textbf{41.18} \\ \bottomrule

\end{tabular}
}
\caption{Localization performance under different \texttt{TraverseGraph} output formats.}
\label{tab:acc_on_different_graph_output}
\end{table*}

% \subsection{Code Graph Construction}\label{sec: append_graph_construct}

% \input{data/api_detail}
\subsection{Tool Output Design}
\subsubsection{Three-level format for \texttt{SearchEntity} output}
Once invoked by the LLM agent, the retrieval APIs search for files, classes, methods, and code snippets in the codebase, and return the results back to the agent. To avoid forming very lengthy code context that may containing noisy information to LLM, we return only necessary information as API outputs. To achieve this, we desgined four granular standard output formats (Figure~\ref{fig:output_format}): \texttt{fold},
\texttt{preview}, \texttt{full code}.

\subsubsection{Tree-based Subgraph Formatting for \texttt{TraverseGraph} Output}\label{sec:study_graph_format}
The \texttt{TraverseGraph} tool traverses the code graph and returns a local subgraph for each input entity. The agent reasons about these subgraphs to understand each entity's complex dependencies. However, reasoning about graphs remains challenging for LLMs. Research by~\cite{fatemi2023talk} demonstrates that LLM performance varies significantly based on graph formatting (how graphs are encoded as text). This makes the format design for output subgraphs crucial.

We have developed a new tree-based format, shown in Figure~\ref{fig:tree_format}, with several features that enhance LLM reasoning: (1) We represent subgraphs as trees, allowing LLMs to use indentation to determine a node's distance from the root, (2) We display complete entity IDs for each node (e.g., \texttt{django/core/validators.py:RegexValidator}) to help LLMs locate nodes easily, and (3) We explicitly specify relation types for each edge, including reversed relations

To evaluate how different graph formats impact code localization performance, we conducted an experiment using 37 challenging samples from SWE-Bench-Lite. These samples were considered "challenging" because they could not be solved by any baseline agent methods. Using \texttt{Claude-3.5} as the Localization Model across all settings, we compared various output formats. Table~\ref{tab:acc_on_different_graph_output} presents our findings. The baseline output formats we tested are described below:

\begin{itemize}
    \item \textbf{row}: For each line, list one row of the adjacency matrix. For example,

    \textit{function "fileA.py:funcA" invokes function "fileA.py:funcB", "fileA.py:funcC"}

    \item \textbf{row (w/ entity attributes)}: Additionally include entity attributes for format \textbf{row}.

    \item \textbf{incident}: The incident format mentioned in~\cite{fatemi2023talk}. An integer instead of entity ID is used to represent each node. For example,

    \textit{Map function "fileA.py:funcA" to index 0. Map function "fileA.py:funcB" to index 1. Map function "fileA.py:funcC" to index 2.}

    \textit{function 0 invokes function 1,2.}

    \item \textbf{Graphviz DOT}: Represent graph in Graphviz DOT language~\cite{ellson2002graphviz}.

    \item \textbf{JSON}: Expand the subgraph as a tree, and convert it to JSON format.
    
\end{itemize}

As shown in Table~\ref{tab:acc_on_different_graph_output}, expanding subgraphs as trees (i.e., \textbf{JSON}, \textbf{tree-based}) can significantly improve the performance. Our \textbf{tree-based} format achieves the best overall performance across different levels of localization tasks. We also test returning entity attributes along with subgraphs. We notice that \textbf{row (w/ entity attributes)} consistently underperforms \textbf{row}, indicating the attributes for all the nodes may be very noisy. Besides, although using incident format can simplify the output and show improvements in file-level localization, it degradation the module- and file-level localization.

\subsection{Implementation}
% \subsubsection{Tool Invocation}
% 描述agent调用tool的过程
To enable the LLM agent to invoke the Code Localization APIs, we handle the interaction differently based on the LLM's capabilities. 
For LLMs that support tool-calling features, we define the tools as a list of JSON objects, which are then used as parameters for the API calls.
For LLMs that do not support tool-calling (such as Qwen), we provide the description of the API and the expected output as part of the LLM's prompt.
When the agent decides to invoke a set of retrieval APIs, it responds with a list of API call names and their corresponding arguments.
These retrieval API requests are processed locally by searching over the built code graph. The results from executing these APIs locally are returned to the agent.

By default, we query the LLM with a temperature setting of 1.0. We conduct two interactions, after which we rerank the results based on mean reciprocal rank (MRR) scores. We also leverage multiprocess execution to speed up the process.
Since all our tools are read-only, \textsc{LocAgent} does not require a specialized Docker environment to operate.
% \subsubsection{Hyperparameters}
% table
% and so on
% \input{6_2_appendix_graph}
\section{Dataset construction and statistics}
\subsection{Dataset construction details}\label{appendix_dataset_construction}
\textbf{Example collection.}
\begin{table*}[ht]
\centering
\resizebox{0.9\textwidth}{!}{
\begin{tabular}{c p{12cm}}
\toprule
\textbf{Category} & \textbf{Keywords} \\ 
\midrule \midrule
\textbf{Performance} & bottleneck, performance improvement, memory usage optimization, time complexity reduction, latency improvement, scalability improvement, CPU usage reduction, caching improvement, concurrency optimization \\ \midrule
\textbf{Security} & Out-of-bounds Write, Out-of-bounds Read, NULL Pointer Dereference, Missing Authorization, memory leak fix, security vulnerability, security issue, authentication bypass, authentication issue, better maintained, buffer overflow, denial of service, security hardening, security patch, unsafe deserialization, Use After Free, Integer Overflow or Wraparound, Uncontrolled Resource Consumption, Missing Authentication for Critical Function \\ 
\bottomrule
\end{tabular}
}
\caption{We use these Keywords to search for Performance and Security related issues with Github Search APIs.}
\label{tab:category_keywords}
\end{table*}

We collected examples on popular Python repositories on Github follow~\cite{jimenez2023swe}. To gather issues related to performance and security, we searched for the keywords listed in Table~\ref{tab:category_keywords} using the GitHub Search APIs. We then used \texttt{GPT-4o-2024-0513} as the classifier based on the issue descriptions.

\noindent \textbf{Ground Truth Locations.}
% 从patch中 parse 出 ground truth
% 清洗数据的说明：document/import/comments 不作为localization目标
The affected files or functions in the original codebase, as identified in the patches, are considered the target locations for the given issue. While it is possible to fix a bug in a location different from the ground truth, the extracted ground-truth locations still serve as approximate targets for localization.
Additionally, edited code such as documents, import statements, and comments are excluded from the localization target. These elements are not considered relevant for bug localization, as they do not directly impact the functionality of the code or its execution. By filtering out these elements, the focus is maintained on the core code changes that are relevant for localization.

% \begin{figure*}[htbp]  % 使用 figure* 环境
%     \centering
%     \includegraphics[width=1\textwidth]{fig/output_locbench.png}
%     \caption{Distribution of Patched Files, Modules, and Functions in Loc-bench.}
%     \label{fig:loc_bench_dist}
% \end{figure*}
% \subsection{Dataset statistics}\label{appendix_dataset_statistics}
% The distribution of modified files, modules, and functions is shown in Figure~\ref{fig:loc_bench_dist}.
% 一些prelimilary的分析：
% 比如说swe-bench中gt file只有一个，gt module的分布， gt function的分布

\section{Additional Experiments}
\begin{table*}[!ht]
    \centering
    \small
    \resizebox{1\textwidth}{!}{
    \begin{tabular}{l|ll ccc cc cc}
        \toprule
        \multirow{2}{*}{\textbf{Type}} & \multirow{2}{*}{\textbf{Method}} & \multirow{2}{*}{\textbf{Loc-Model}} & \multicolumn{3}{c}{\textbf{File} (\%)} & \multicolumn{2}{c}{\textbf{Module} (\%)} & \multicolumn{2}{c}{\textbf{Function} (\%)}\\
        \cmidrule(lr){4-6} \cmidrule(lr){7-8} \cmidrule(lr){9-10}
         & & & \textbf{NDCG@1} & \textbf{NDCG@3} & \textbf{NDCG@5} & \textbf{NDCG@5} & \textbf{NDCG@10} & \textbf{NDCG@5} & \textbf{NDCG@10} \\
        \midrule
        \midrule
        \multirow{5}{*}{Embedding-Based} 
            & \multicolumn{2}{l}{BM25~\cite{robertson2009probabilistic}} & 38.69	&46.5	&50.61	&37.31	&39.86	&26.15	&27.92 \\
            & \multicolumn{2}{l}{E5-base-v2~\cite{wang2022text}} & 49.64	&64.19	&66.6	&53.15	&54.45	&31.39	&35.3\\
            % & \multicolumn{2}{l}{CodeT5+~\cite{wang2023codet5opencodelarge}} & 34.67 & 56.93 & 67.15 & 50.73 & 58.39 & 25.55 & 32.85\\
            & \multicolumn{2}{l}{Jina-Code-v2~\cite{günther2023jinaembeddingsnovelset}} & 43.43	&59.93	&63.7	&51.02	&54.13	&33.28	&36.44\\
            & \multicolumn{2}{l}{Codesage-large-v2~\cite{zhang2024code}}  & 47.81	&60.82	&64.39	&49.38	&52.22	&27.03	&30.74\\
            & \multicolumn{2}{l}{CodeRankEmbed~\cite{suresh2024cornstack}} & 52.55	&67.54	&70.39	&57.51	&59.76	&40.28	&42.55\\
        \midrule
        \multirow{2}{*}{Procedure-Based} 
            & \multirow{2}{*}{\makecell[l]{Agentless\\\cite{xia2024agentless}}}  
                & \texttt{GPT-4o}   & 67.15	&71.76	&71.76	&64.31	&64.31	&53.81	&53.81 \\
            & & \texttt{Claude-3.5} & 72.63	&76.72	&76.87	&67.36	&67.36	&57.55	&57.55 \\
        \midrule
        \multirow{9}{*}{Agent-Based} 
            & \multirow{2}{*}{\makecell[l]{MoatlessTools\\\cite{orwall2023moatless}}} 
                & \texttt{GPT-4o}   & 73.36	&80.03	&80.33	&68.57	&69.09	&49.77	&50.62  \\
            & & \texttt{Claude-3.5} & 72.63	&80.73	&80.88	&69.11	&69.11	&53.03	&53.16  \\
        \cmidrule(lr){2-10}
            & \multirow{2}{*}{\makecell[l]{SWE-agent\\\cite{yang2024sweagent}}}
                & \texttt{GPT-4o}   & 57.3	&63.96	&64.12	&53.95	&53.95	&42.32	&42.44 \\
            &                      & \texttt{Claude-3.5} & 77.37	&84.32	&84.93	&72.77	&72.9	&59.67	&59.79 \\
        \cmidrule(lr){2-10}
            & \multirow{2}{*}{\makecell[l]{Openhands\\\cite{wang2024openhands}}} 
                & \texttt{GPT-4o}   & 60.95	&67.62	&68.39	&58.18	&58.6	&44.34	&44.66  \\
            & & \texttt{Claude-3.5} & 76.28	&84.27	&84.43	&75.79	&75.92	&63.13	&63.8  \\
        \cmidrule(lr){2-10}
            & \multirow{3}{*}{LocAgent (Ours)}
                % & \texttt{GPT-4o}      &63.14	&84.31	&85.77	&74.45	&77.37	&59.85	&65.33  \\
            & \texttt{Qwen2.5-7B(ft)}      & 70.80 & 79.36 & 80.9 & 70.99 & 71.68 & 55.62 & 58.09  \\
            & & \cellcolor{blue!10}\texttt{Qwen2.5-32B(ft)} & \cellcolor{blue!10}75.91 & \cellcolor{blue!10}84.74 & \cellcolor{blue!10}85.64 & \cellcolor{blue!10}76.28 & \cellcolor{blue!10}76.77 & \cellcolor{blue!10}64.27 & \cellcolor{blue!10}65.93  \\
            & & \cellcolor{blue!20}\texttt{Claude-3.5}   & \cellcolor{blue!20}\textbf{77.74} & \cellcolor{blue!20}\textbf{86.19} & \cellcolor{blue!20}\textbf{87.14} & \cellcolor{blue!20}\textbf{77.73} & 
            \cellcolor{blue!20}\textbf{78.1} & 
            \cellcolor{blue!20}\textbf{64.34} & \cellcolor{blue!20}\textbf{65.57} \\
        \bottomrule
    \end{tabular}
    }
    \caption{NDCG scores comparison showing ranking quality of different methods.}
    % \label{tab:model_performance}
    \label{tab:ndcg_comparison}
\end{table*}

\subsection{Implementation Details}
\subsubsection{Baselines Implementation}\label{appendix_exp_impl}

Regarding the embedding-based methods in our evaluation, these approaches operate primarily at the function level, where each function is embedded as a separate unit. The function's context (its containing file and class) is appended to the function representation before embedding, rather than being embedded separately. While theoretically these methods could employ hierarchical indexing, the standard implementations we evaluated use flat indexing structures where each function is embedded as a single unit.

We use OpenHands’s remote runtime feature to parallelize evaluation on OpenHands and SWE-agent. We use Openhands version 0.12.0 released on Oct 31, 2024.

\subsubsection{Quantifying Task Difficulty Based on Code Graph Distance}\label{appendix_diff_dist}
We measure task difficulty by computing the average shortest hop distance between the functions mentioned in the issue descriptions and the patched functions within our code graph. Specifically, we first extract potential function names from each issue description using \texttt{GPT-4o-2024-0513}, and identify their corresponding nodes in the code graph using the global dictionary. These identified nodes form the set of predicted nodes, denoted as $\mathcal{C}$. Similarly, we link the ground truth functions from the patch to their corresponding nodes in the code graph, forming the set of target nodes, denoted as $\mathcal{T}$.
To quantify the difficulty $\delta$, we calculate the average shortest hop distance between the predicted nodes $\mathcal{C}$ and the target nodes $\mathcal{T}$, defined as:
\[
\delta =\frac{1}{|\mathcal{C}|}\sum_{c\in \mathcal{C}}\frac{1}{{min}_{t\in \mathcal{T}} d(c, t)+1}
\]
where $d(c, t)$ represents the shortest hop distance between nodes $c$ and $t$ in the graph.
For performance analysis stratified by difficulty, we round $\delta$ down to $\lfloor \delta \rfloor$ to group samples by difficulty levels, and we exclude samples where the LLM fails to extract any valid function names.

\subsubsection{Training details.}\label{appendix_ft_detail}
\textbf{Fine-tuning Settings.} 
We use \texttt{Qwen-2.5-Coder-Instruct}~\cite{hui2024qwen25codertechnicalreport} 7B and 32B variants as our base models. We fine-tuned \texttt{Qwen-2.5-Coder-Instruct} 7B and 32B models on 768 training samples from the SWE-Bench training dataset, leveraging LoRA for efficient adaptation. The training set included 447 samples generated by \texttt{Claude-3.5}, while the remaining samples were iteratively generated using the fine-tuned \texttt{Qwen2.5-32B} model. The fine-tuning process was conducted over 5 epochs with \textit{max\_token} set to 128\textit{k} and a learning rate of $2 \times 10^{-4}$.

% \textbf{Metrics}
% Let \( \text{GT}_i \) represent the set of ground truth locations of the specific level for instance \( i \), \( \text{Loc}_k(i) \) represent the top \( k \) localization results for instance \( i \). 
% Thus, the recall can be expressed as:
% \[
% Acc@k = \frac{\sum_{i=1}^{N} \mathbf{1}(\text{Loc}_k(i) \supseteq \text{GT}_i)}{N}
% \]
% \noindent Where \( N \) is the total number of instances,  \( \supseteq \) denotes a superset, meaning all locations in \( \text{GT}_i \) are found within the top \( k \) results, and \( \mathbf{1}(\cdot) \) is an indicator function that returns 1 if the condition inside is true (i.e., if the top \( k \) localization results for instance \( i \) are a superset of the ground truth) and 0 otherwise.
\section{Prompt}
\label{sec:appendix_prompt}
In this section, we go through the prompt template that make up the agent’s history.
% , discussing them in the order of presentation to SWE-agent.
\begin{figure*}[hb!]
  \centering
  \includegraphics[width=\textwidth, trim=39 93 355 55, clip]{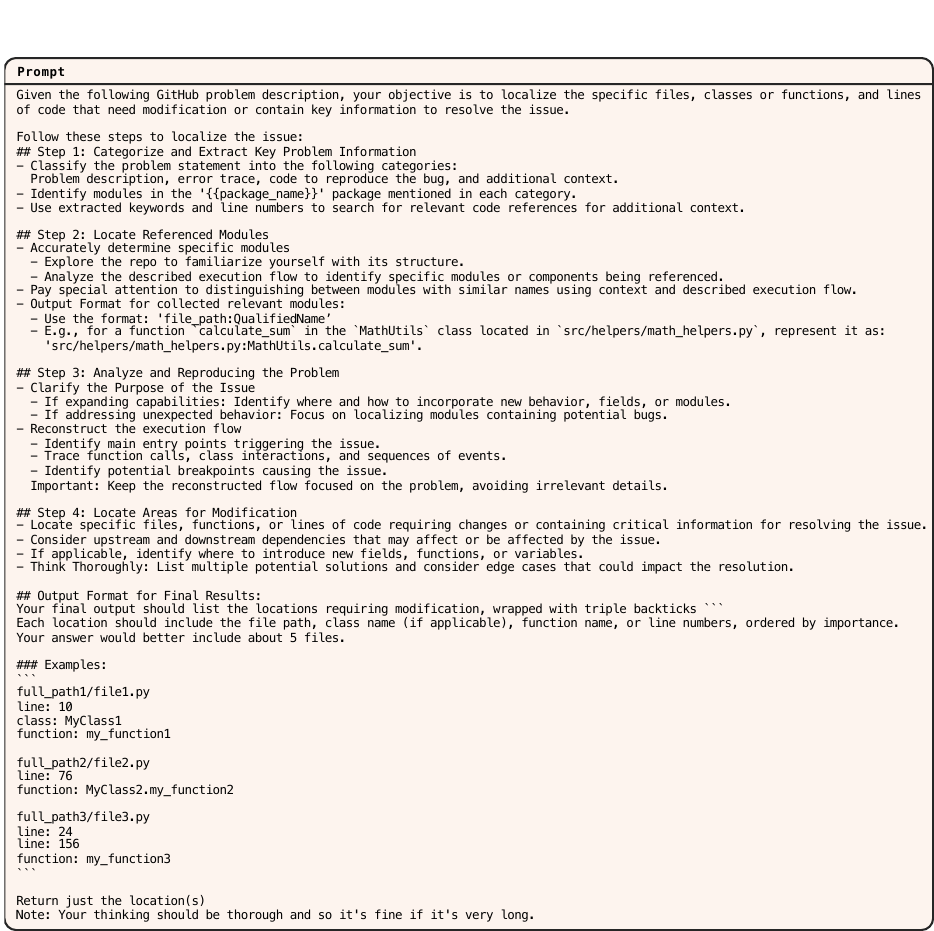}
  \caption{The task instruction prompt for \textsc{LocAgent}.}
\end{figure*}

\end{document}